\documentclass[12pt,twoside,epsfig]{report}

\advance\topmargin by -1.6cm

\newenvironment{eq}[1]
{\[\begin{array}{#1}}{\end{array}\]}

\let\rvec=\vec        

\def\plint {\hskip-2mm{^\pl\hskip-3mm\int}}

\def\lq{\hskip-0.5mm\setminus\hskip-0.5mm} 
  
 \def\({\Bigl(}
\def\){\Bigr)}   
 \def\|{\Big|}
\def\then{\Rightarrow}  
 \def\o{\circ}
\def\m{\bullet}    
\def\x{\times}
   
\def\ox{\otimes}

\def\pl{{~\oplus~}}

\def\PL{\displaystyle \bigoplus}
\def\SUM{\displaystyle \sum}

\def\mid{\big\bracevert}

\def\sub{\subseteq}
\def\subnoteq{\subset}
\def\sup{\supseteq}
\def\supnoteq{\supset}
\def\and{\wedge}
\def\And{\bigwedge}
\def\AND{\displaystyle\bigwedge}
\def\od{\vee}

\def\OD{\displaystyle\bigvee}

\def\rin{{\,\in\kern-.42em\in}}

\def\spec{\,{\rm spec}\,}
\def\rep{\,{\rm rep}\,}

\def\tr{{\,{\rm tr }\,}}
\def\det{\,{\rm det }\,}

\def\CLIFF{\,\hbox{CLIFF}}

\def\Int{\,{\rm Int}\,}

\def\card{\ro{\,card\,}}

\def\sx{~\rvec\x~\!}

\def\irrep{{{\bf irrep\,}}}
\def\rep{{{\bf rep\,}}}

\def\A{{\,{\rm A\kern-.55emA}}}
\def\B{{\,{\rm I\kern-.2emB}}}
\def\C{{\,{\rm I\kern-.55emC}}}
\def\E{{\,{\rm I\kern-.2emE}}}
\def\G{{\,{\rm I\kern-.55emG}}}
\def\H{{{\rm I\kern-.2emH}}}
\def\I{{\,{\rm I\kern-.2emI}}}
\def\K{{\,{\rm I\kern-.2emK}}}
\def\L{{\,{\rm I\kern-.2emL}}}
\def\M{{\,{\rm I\kern-.16emM}}}
\def\N{{\,{\rm I\kern-.16emN}}}
\def\Q{{\,{\rm I\kern-.5emQ}}}
\def\R{{{\rm I\kern-.2emR}}}
\def\S{{\,{\rm I\kern-.42emS}}}
\def\T{{\,{\rm I\kern-.37emT}}}
\def\UU{{\,{\rm I\kern-.51emU}}}
\def\Z{{\,{\rm Z\kern-.32emZ}}}

\def\p{\partial}

\def\al{\alpha}  \def\be{\beta} \def\ga{\gamma}
\def\de{\delta}  \def\ep{\epsilon}  
\def\th{\theta}   \def\vth{\vartheta} 
\def\ka{\kappa}   \def\la{\lambda}   \def\si{\sigma}
\def\De{\Delta}   \def\om{\omega} \def\Om{\Omega}
\def\phi{\varphi} 
 \def\Ga{\Gamma}  
    \def\La{\Lambda}

\def\vec#1{\underline{\bf vec}_{#1}}

\def\GL{{\bf GL}}  
\def\SL{{\bf SL}}
\def\U{{\bf U}} 
 
\def\O{{\bf O}}   
\def\SU{{\bf SU}} 
 
\def\SO{{\bf SO}}

 \def\D{{\bl D}}

\def\d#1{{\check{#1}}}
\def\angle#1{\langle#1\rangle}

\def\rstate#1{|#1\rangle}

\def\brack#1{\lbrack#1\rbrack}

\def\ro#1{{\rm #1}}
\def\bl#1{{\bf {#1}}}
\def\cl#1{{\cal #1}}

\def\ol#1{\overline{#1}}

\def\dprod#1#2{\langle#1,#2\rangle}
\def\sprod#1#2{\langle#1|#2\rangle}

\def\com#1#2{\lbrack#1,#2\rbrack}

\def\acom#1#2{\{#1,#2\}}

\def\map{\longrightarrow}
\def\inmap{\hookrightarrow}
\def\lrmap{\leftrightarrow}

\def\dmap{\Big\downarrow}

\def\mape{\longmapsto}

\def\Diagr#1#2#3#4#5#6#7#8{\matrix{\noalign{\vskip5mm}
      \hskip-4mm&              &{\scriptstyle #5}&              &     \cr
      \hskip-4mm& #1           & \map           & #2           &     \cr
{\scriptstyle #8}\hskip-3mm   &\dmap         &    &\dmap&\hskip-4mm{\scriptstyle#6} \cr
      \hskip-4mm& #4           & \map           & #3           &     \cr
      \hskip-4mm&              &{\scriptstyle#7}&              &     \cr
\noalign{\vskip5mm}             }}

\def\plintq3{{}^\pl\hskip-2.5mm\int_{\R^3}{d^3q\over 2q_0(2\pi)^3}} 
 
\begin{document}

\begin{titlepage} 
 
$~$
\hfill MPP-2003-135
\vskip25mm
\centerline{\bf  HARMONIC ANALYSIS OF SPACETIME}
\vskip2mm
 \centerline{\bf WITH HYPERCHARGE AND ISOSPIN}\vskip1cm

\centerline{
Heinrich Saller\footnote{\scriptsize hns@mppmu.mpg.de} }
\centerline{Max-Planck-Institut f\"ur Physik}
\centerline{Werner-Heisenberg-Institut}
\centerline{M\"unchen, Germany}

\vskip25mm

\centerline{\bf Abstract}
\vskip5mm
In analogy to the harmonic analysis for the Poincar\'e group with
its irreducible representations characterizing free particles,
the harmonic analysis for a nonlinear spacetime model as homogeneous space of the
extended Lorentz group $\GL(\C^2)$ is given. What the Dirac energy-momentum
measures are for free particles, are multipole  measures in the 
analysis of nonlinear spacetime - they are related to spacetime interactions. The representations induced from
the nonlinear spacetime fixgroup $\U(2)$ connect representations of
external (spacetime-like) degrees of freedom with those of internal
(hypercharge-isospinlike) ones as seen in the standard model of electroweak
and strong interactions.
The methods used are introduced and exemplified with the
 nonrelativistic Kepler
dynamics in an interpretation as harmonic analysis of time and position
functions. 

\vskip1cm

\end{titlepage}

{\small \tableofcontents} 

\newpage

\chapter{Orbits of Time and Space}

This chapter serves as a - very long - introduction.
It treats with the  nonrelativistic Kepler dynamics - classical and quantum.
The Kepler dynamics with the ${1\over r}$-potential   has, on the one side,
 enough
structure and, on the other side,   is familiar enough 
to practice the 
language which will be talked  in the
following  chapters  for 
particles and interactions, related to relativistic 
 spacetime. It  exemplifies the necessary  concepts - 
operational sym\-met\-ries and  orbits, realizations and
re\-pre\-sen\-ta\-tions\cite{FULHAR,GIL,HEL1,KIR} of, especially, time and position, 
compact and noncompact for  spherical, flat and hyperbolic spaces, scattering and bound states,
special functions as matrix elements\cite{VIL}, 
homogeneous spaces, invariant measures,
multipole measures,  harmonic expansions
etc.

\section{Symmetries of the Kepler Dynamics}

The  Kepler Hamiltonian 
\begin{eq}{l}
H={\rvec p^2\over 2}+{\de\over r},~~\de=\pm1 ~\hbox{ (repulsion, attraction)}\cr
\end{eq}has, in addition to rotation invariance with conserved angular momentum 
$\rvec {\cl L}$,
the {Lenz-Runge invariance} with
conserved {Lenz-Runge vector} $\rvec{\cl P}$ 
(for the sun system called perihelion vector and perihelion conservation - no
rosettes) given with the  Poisson Lie-bracket
$[f,g]_P=
{\p f\over \p\rvec p}
{\p g\over \p\rvec x}
-{\p f\over \p\rvec p}
{\p f\over \p\rvec x}$ from $[p^a,x^b]_P=\de^{ab}$
\begin{eq}{l}
\rvec{\cl L}=\rvec x\x\rvec p,~~
\rvec{\cl P}=\rvec p\x\rvec {\cl L}+\de {\rvec x\over r}
\then\left\{\begin{array}{rl}
\com H{\rvec {\cl P}}_P&=0\cr
\com{\cl L^a}{\cl P^b}_P&=-\ep^{abc}\cl P^c\cr
\com{\cl P^a}{\cl P^b}_P&=2H\ep^{abc}\cl L^c\cr\end{array}\right.
\end{eq}

The Kepler Hamiltonian should be compared in the following 
with the free Hamiltonian (constant potential)
\begin{eq}{l}
H_0={\rvec p^2\over 2}+V_0
\then\left\{\begin{array}{rl}
\com {H_0}{\rvec p}_P&=0\cr
\com{\cl L^a}{p^b}_P&=-\ep^{abc}p^c\cr
\com{p^a}{p^b}_P&=0\cr\end{array}\right.
\end{eq}As seen below, the Lenz-Runge vector
$\rvec{\cl P}$ expands the translations $\rvec p$.

With the Hamiltonian in the Lenz-Runge vector brackets
three different types for the  energy values 
from the spectrum of the Kepler Hamiltonian
have to be distinguished 
to classify its possible symmetries.
In all three cases 
$E=0$, $E>0$ and $E<0$,
the dynamics is characterized by a  
real 6-di\-men\-sio\-nal invariance Lie algebra of rank 2, i.e. with two 
independent invariants.

For trivial energy, the symmetry
 is - as for the free Hamiltonian -
 the semidirect Euclidean structure
with the  rotations and translations 
in three di\-men\-sions\footnote{\scriptsize
$\log G$ denotes the Lie algebra of a Lie group $G$, e.g.
$\log \SU(2)=A_1^c$. $G\sx H$ denotes a semidirect product with the group $G$
acting upon the group $H$.}
\begin{eq}{rl}
\spec H\ni E=0:&\left\{\begin{array}{l}
\com{l^a}{l^b}=-\ep^{abc}l^c,~\com{l^a}{p^b}=-\ep^{abc}p^c,~
\com{p^a}{p^b}=0\cr
\hbox{invariants: }
\rvec p^2,~~ \rvec l\rvec p\cr
\hbox{re\-pre\-sen\-ta\-tion: }l^a\mape \cl L^a,~~p^a\mape \cl P^a\cr
\end{array}\right.\cr
\hbox{Lie algebra:}&A_1^c\rvec \pl \R^3=\log[\SO(3)\sx \R^3]\cong\R^6\cr

\end{eq}The angular momentum invariant $\rvec l^2$ is no translation invariant.

A nontrivial energy can be  used to renormalize
the Lenz-Runge vector 
\begin{eq}{l}
\begin{array}{r}
\spec H\ni E\ne 0,~~\rvec{\cl B}={\rvec{\cl P}\over\sqrt{2|H|}}:\cr
\ep(E)={E\over|E|}={H\over|H|}~~~~\cr
\end{array}

\left\{\begin{array}{rl}
\com{\cl L^a}{\cl L^b}_P
&=-\ep^{abc}\cl L^c\cr
\com{\cl L^a}{\cl B^b}_P&=-\ep^{abc}\cl B^c\cr
\com{\cl B^a}{\cl B^b}_P&=\ep(E)\ep^{abc}\cl L^c\cr
\end{array}\right.
\end{eq}

Positive energies lead to scattering orbits
(classical and quantal), there arises
the Lie algebra of
the  noncompact Lorentz group
  with the Lenz-Runge vector defining the
`boosts' (called so in analogy - no special relativistic transformations)  
\begin{eq}{rl}
E>0:&\left\{\begin{array}{l}
\com{l^a}{l^b}=-\ep^{abc}l^c=-\com{b^a}{b^b},~~\com{l^a}{b^b}=-\ep^{abc}b^c\cr
\hbox{invariants: }
\rvec {l}^2-\rvec {b}^2,~\rvec {l}\rvec{b}\cr
\hbox{re\-pre\-sen\-ta\-tion: }l^a\mape \cl L^a,~~b^a\mape \cl B^a\cr

\end{array}\right.\cr
\hbox{Lie algebra:}&A_1^c\pl i A_1^c\cong\log\SO(1,3)\cong\R^6\cr
\end{eq}

For negative energies, leading to bound orbits (classical and quantal), the symmetries constitute 
the compact Lie algebra of $\SO(4)$,
 locally isomorphic to $\SO(3)\x\SO(3)$
\begin{eq}{rl}
E<0:&\left\{\begin{array}{l}
\com{l^a}{l^b}=-\ep^{abc}l^c=\com{m^a}{m^b},~~\com{l^a}{m^b}=-\ep^{abc}m^c\cr
\rvec l_\pm={\rvec l\pm\rvec m\over2}\then
\com{l_\pm^a}{l_\pm^b}=-\ep^{abc}l_\pm^c,~
\com{l_+^a}{l_-^b}=0\cr
\hbox{invariants: }\rvec l^2+\rvec m^2
=2(\rvec l_+^2+\rvec l_-^2),~
\rvec {l}\rvec{m}=\rvec{l}_+^2-\rvec{l}_-^2\cr
\hbox{re\-pre\-sen\-ta\-tion: }l^a\mape \cl L^a,~~m^a\mape\cl B^a\cr
\cr
\end{array}\right.\cr
\hbox{Lie algebra:}
&A_1^c\pl A_1^c\cong\log[\SO(3)\x \SO(3)]\cong\R^6\cr\end{eq}

The transition from positive to negative energies can be formulated as a
transition from real momenta to imaginary `momenta'
\begin{eq}{l}
\sqrt{2E}=\left\{\begin{array}{rll}
|\rvec q|,&E>0,&\hbox{scattering orbits, real momenta}\cr
i| Q|,&E<0,&\hbox{bound orbits, imaginary `momenta'}\cr\end{array}\right.
\end{eq}

The three {Kepler Lie groups}
can be decomposed - as manifolds - 
with  rotation group  classes\footnote{\scriptsize
The subindex in $\SO_0(1,s)$ denotes the unit connection component.}
\begin{eq}{rl}
\SO(4)&\cong\SO(3)\x \SO(4)/\SO(3)\cr
\SO_0(1,3)&\cong\SO(3)\x\SO_0(1,3)/\SO(3)\cr
\end{eq}The rotation group
itself is a product of 
 axial rotations $\SO(2)$ with  the 2-sphere\footnote
{\scriptsize
The $s$-sphere and the $s$-hyperboloid are denoted by
$\Om^s\cong\SO(1+s)/\SO(s)$ and 
$\cl Y^s\cong\SO_0(1,s)/\SO(s)$ resp. for $s=1,2,\dots$
with $\cl Y^{1+s}\cong\cl Y^1\x\Om^s$.}
$\Om^2\cong\SO(3)/\SO(2)$
 for the angular momentum direction, i.e. $\Om^2$ is the orientation manifold
 of the  axial rotations. 
The Lenz-Runge operations, compact as  3-sphere 
$\Om^3=\SO(4)/\SO(3)$ and noncompact  
 as 3-hyperboloid $\cl Y^3=\SO_0(1,3)/\SO(3)$, are
the orientation manifolds - or equivalence classes -  of the rotation group.
Also these symmetric spaces have a manifold decomposition
into a characteristic Abelian subgroup 
and a 2-sphere for the Lenz-Runge vector direction
\begin{eq}{l}
{\scriptsize\pmatrix{
\hbox{spherical}\cr
\hbox{flat, Euclidean}\cr
\hbox{hyperbolic}\cr}}:~~

{\scriptsize\pmatrix{
\SO(4)\cr
\SO(3)\sx\R^3\cr
\SO_0(1,3)\cr}}
\cong \SO(2)\x\Om^2\x
{\scriptsize\pmatrix{
\SO(2)\cr
\R\cr
\SO_0(1,1)\cr}}\x\Om^2
\end{eq}The 2-di\-men\-sional Abelian subgroups  reflect the 
rank 2 with the two independent invariants.
The  rotation group $\SO(3)$ in all groups determines the 
angular momentum.
The re\-pre\-sen\-ta\-tion of the 
2nd Abelian factor
decides on the spherical, parabolic (or flat) and hyperbolic orbits.

\section{Classical Time Orbits}

In classical theories one is primarily interested 
in the time orbits\footnote{\scriptsize
A group $G$, realized by bijections (permutations, automorphisms)
on a set $S$, defines an orbit $G\m x\sub S$ for each element $x\in S$
and a fixgroup $G_x\sub G$ with $G\m x\cong G/G_x$. The group action decomposes
the set
into disjoint orbits $G\m x\in S/G$. For a vector space $S=V$, the orbit
displays matrix elements $\dprod \om{g\m x}$
of the group representation $G\ni g\mape \GL(V)$.}
the mass points perform
 in position space
$\R\ni t\mape \rvec x(t)\in\R^3$, mathematically 
in the irreducible  realizations of the time translation group $\R$.
The characterizing eigenvalues, i.e. the invariant energies,
are classically imposed by boundary or initial conditions.

\subsection{Time  Orbits in Position Space}

The functions of position and momentum $(\rvec x,\rvec p)$
 build
 an associative unital algebra, for a classical framework 
commutative with the pointwise product.  
The problems with the Kepler potential ${1\over r}$ 
at the origin $\rvec x=0$ are
neglected, they  deserve a more careful discussion. 
This algebra has a noncommutative Lie algebra structure
with the Poisson bracket. Therewith the 
three Kepler Lie algebras $\log\SO(4)$, $\log\SO_0(1,3)$ and $\log[\SO(3)\sx\R^3]$
act adjointly on this algebra.

The products of angular momentum 
with position, momentum and Lenz-Runge vector
vanish (orthogonality)
\begin{eq}{l}
\rvec {\cl L}\rvec x=0,~~
\rvec {\cl L}\rvec p=0,~~
\rvec {\cl L}\rvec{\cl P}=0
\end{eq}i.e., position
$\rvec x$, momentum $\rvec p$ and Lenz-Runge vector $\rvec{\cl P}$
are in the position orbit plane.
For  gravity in the sun system the  orbit planarity following from 
  rotation invariance  constitutes  Kepler's 1st law. 

The invariant squares  of angular momentum
and perihelion vector combine the
Hamiltonian and determine the  energy $E$ 
by angular momentum value $L$ and perihelion value $P$
\begin{eq}{rl}
\rvec {\cl P}^2&=1+2H\rvec {\cl L}^2\then
 E= {P^2-1\over 2L^2}\hbox{ with }|\rvec {\cl P}|=P,~|\rvec {\cl L}|=L\cr
-{1\over 2H}&=\rvec {\cl L}^2-\ep(E)\rvec{\cl B}^2
\end{eq}

The time orbits in position space are conic sections, described
by polar equations
with one focus as origin (2nd Kepler law) 
\begin{eq}{rl}
\rvec{\cl P}\rvec x=P r \cos\phi&=
(\rvec p\x\rvec {\cl L})\rvec x+\de  r =
L^2+\de  r \cr
\then
r(\phi)&={L^2\over  P\cos\phi-\de }\hbox{ with }\de=\pm1
\end{eq}$\rvec x$ shows to the
peri- and aphelion for $\phi=0$ and $\phi=\pi$ resp.
The invariants can be expressed by perihelion distance $r_0$ and 
momentum $p_0$ 
as possible initial conditions 
\begin{eq}{l}
\hbox{for }\phi=0:~~ \rvec x_0\rvec p_0=0\then\left\{\begin{array}{rl}
L&=r_0p_0\cr
P&=r_0p_0^2+\de\cr
E&={p_0^2\over 2}+{\de\over r_0}\cr
\end{array}\right.
\end{eq}

The connection between  Cartesian 
$\rvec x=(x,y,0)$ and 
and  polar  equations
is given in the following table

{\scriptsize
\begin{eq}{c}
\begin{array}{|c||c|c|c|}\hline
\hbox{energy }E={P^2-1\over 2L^2}
&E<0&E>0&E=0\cr\hline\hline
\hbox{group orbit}
&\SO(2),\hbox{ ellipse}&\I(2)\x\SO_0(1,1),\hbox{ hyperbola}&
\R,\hbox{ parabola}\cr\hline
\hbox{Cartesian equation}&
\begin{array}{c}
{x^2\over a^2}+{y^2\over b^2}=1\cr a\ge b\end{array} 
&\begin{array}{c}
{x^2\over a^2}-{y^2\over b^2}=1\cr\hbox{(right and left branch)}\end{array}
&\begin{array}{c}y^2=-2d x\cr d>0\end{array}\cr\hline
\hbox{foci distance }2c&c^2=a^2-b^2&c^2=a^2+b^2&
\hbox{(one focus)}\cr\hline
\begin{array}{c}
\hbox{polar equation}\cr
\hbox{(with pole}\cr
\hbox{in the right focus)}\end{array}
&\begin{array}{c}
r(\phi)={L^2\over P\cos\phi+1}\cr
0<P<1\end{array}
&\begin{array}{c}
r_{R,L}(\phi)={L^2\over P\cos\phi\pm 1}\cr
P>1\end{array}
&\begin{array}{c}
r(\phi)={L^2\over \cos\phi+1}\cr
P=1\end{array}\cr\hline
\begin{array}{c}
\hbox{distance}\cr
\hbox{of pole}\cr
\hbox{to peri- and aphelion}\end{array}
&\begin{array}{c}
(a-c,a+c)\cr=(r(0),r(\pi))\cr=({L^2\over1+P},{L^2\over1-P})
\end{array}
&\begin{array}{c}
(c-a,c+a)\cr
=(r_R(0),r_L(0))\cr
=({L^2\over P+1},
{L^2\over P-1})\end{array}
&\begin{array}{c}
{d\over2}\cr
=r(0)\cr
={ L^2\over2}\end{array}
\cr\hline

(a,c,b)=&
(-{1\over 2E},-{P\over 2E},{L\over \sqrt{-2E}})&
({1\over 2E},{P\over 2E},{L\over \sqrt{2E}})&\cr\hline
(L^2,P)=&({b^2\over a},{c\over a})&({b^2\over a},{c\over a})&(d,1)
\cr\hline
\end{array}\cr\cr
\hbox{\bf  Cartesian and Polar  Equations for Conic Sections}
\end{eq}
}

The  squared length   of  the perihelion vector is
the discriminant of the 2nd order polynomial which occurs  in the equation of motion 
\begin{eq}{rl}
{dr\over dt}=[H,r]_P=p_r,~~p_r^2&={2Er^2-2\de r-L^2\over r^2}\cr
-\det{\scriptsize\pmatrix{2E&\de\cr\de&-L^2\cr}}&=1+2EL^2=P^2\hbox{ for }\de^2=1
\end{eq}

{  Attraction} 
$\de=-1$, e.g.
in gravity or with  charge
numbers of opposite sign $z_1z_2<0 $ in electrostatics, can come with
negative and positive energies for compact and noncompact orbits resp.
\begin{eq}{l}
\de =-1:~
P^2=1+2E L^2
\left\{\begin{array}{ll}
<1\iff&-{1\over 2 L^2}\le E<0\cr
&\hbox{(ellipse)}\cr
=1\iff& E=0\cr
&\hbox{(parabola)}\cr
>1\iff& E>0\cr
&\hbox{(hyperbola branch  around pole)}\cr\end{array}\right.\cr
\end{eq}Repulsion
$\de =1$, e.g. for charge numbers of equal sign
$z_1z_2>0$, has positive energies only (noncompact orbits)
\begin{eq}{ll}
\de=1:~
P^2=1+2E L^2>1\iff&E>0\cr
&\hbox{(hyperbola branch, not around pole)}\cr
\end{eq}

Since the angular momentum is twice the time change of the  orbit area
\begin{eq}{l}
\rvec{\cl L}=\rvec x\x d_t\rvec x= 2d_t\rvec {\cl A}
\end{eq}the orbit area $A=|\rvec{\cl  A|}$ and the orbit time $T$ 
for ellipses are related to the
conserved angular momentum
 which relates the large axis of the 
ellipse to the orbit time (Kepler's 3rd law)
\begin{eq}{l}
L=2{A\over  T}=2{\pi ab\over T}
\then a^3=({T\over 2\pi})^2=-{1\over 8E^3}
\end{eq}

For the free theory the orbits are lines
\begin{eq}{l}
\de=0:~~E={\rvec p^2\over 2},~~r(\phi)={r(0)\over \cos\phi}~~
{\scriptsize\pmatrix{x\cr y\cr}}=
{\scriptsize\pmatrix{1&0\cr\xi&1\cr}}{\scriptsize\pmatrix{1\cr0\cr}}
\end{eq}

\subsection{Orbits as  Time Classes}

A  dynamics leads to realizations of the time translation group.
The solutions in  classical physics
are irreducible
time orbits in position space.
They have to be isomorphic to quotient groups of time $\R$
with the kernel $K$ of the time realization
\begin{eq}{l}
\R\ni t\mape \rvec x(t)\in\R^3,~~\rvec x[\R]\cong \R/K\subnoteq\R^3
\end{eq}The 
$\R$-subgroups which can arise as kernels for Lie quotient groups are
the  full group $\R$, the  trivial group $\{0\}$ and - up to isomorphy - the
discrete group $\Z$.

All possible quotient groups $\R/K$ (time  equivalence classes)
for time realizations  are
seen in the sky 
as solutions of the Kepler dynamics. 
The trivial time re\-pre\-sen\-ta\-tion is seen in 
our  sun, assumed infinitely heavy, with trivial orbit $\R/\R\cong\{1\}$
and energy $E=-\infty$. 
The  compact unfaithful time realizations with kernel $\Z$  
give  bound orbits (planets on ellipses), isomorphic to 
the torus $\U(1)$. For the earth, $\Z$ counts the years.
The  noncompact faithful realizations, isomorphic to full $\R$,  
give  scattering orbits (never returning comets on one branch
 of an hyperbola or on a parabola). 

The conic sections (ellipses and hyperbolas) for the planar orbits 
have the metrical tensors 
\begin{eq}{l}
{x^2\over a^2}\pm {y^2\over b^2}=1,~
{\scriptsize\pmatrix{{1\over a^2}&0\cr 0&\pm {1\over b^2}\cr}}=
{\scriptsize\pmatrix{4E^2&0\cr 0&- {2E\over L^2}\cr}}=
{4E^2\over L\sqrt{2|E|}}  {\scriptsize\pmatrix{\eta &0\cr
0&\pm{1\over \eta }\cr}}\cr
\end{eq}with
the  ratio of the units
for the two directions $\eta={b\over a}=L\sqrt{2|E|}$ 
the product of  angular momentum with energy.
The  orbits in position space can be  parametrized as follows 
 \begin{eq}{lrl}
\hbox{ellipses:}&\SO(2)\cong\R/\Z:&
{\scriptsize\pmatrix{ x\cr y\cr}}=
{\scriptsize\pmatrix{
\cos\chi& -\eta \sin\chi\cr
{1\over \eta }\sin\chi& \cos\chi\cr}}
{\scriptsize\pmatrix{ 1\cr 0\cr}}\cr
\hbox{hyperbolas:}&\SO_0(1,1)\cong\R:&{\scriptsize\pmatrix{ x\cr y\cr}}=
{\scriptsize\pmatrix{
\cosh\psi& \eta \sinh\psi\cr
{1\over \eta }\sinh\psi& \cosh\psi\cr}}
{\scriptsize\pmatrix{\pm 1\cr 0\cr}}\cr
\end{eq}The time parametrization comes via
time dependent group parameters, $t\mape\chi(t),\psi(t)$.

How are the time orbits, i.e. the quotient groups $\R/K$,  related to 
the full Kepler groups  $\SO(4)$, $\SO_0(1,3)$ and $\SO(3)\sx\R^3$ 
acting upon the algebra with the position-momentum 
functions? An individual solution  of a dynamics (time realizations), 
here one orbit $\rvec x$ in position space,  has not to have
all the invariances $G$ of the Hamiltonian.
$\rvec x$-equivalent solutions are on the orbit
$G\m \rvec x\cong G/H$ of the invariance group
$G$ in the solution space (not in position space) 
with a remaining subgroup $H$-symmetry. 
E.g.,
by  rotating the initial conditions of one solution for a rotation 
invariant Hamiltonian one obtains an equivalent solution.
The fixgroup $H$ of a solution is the
$G$-subgroup which leaves the  orbit 
in position space invariant  - as a whole, not its
individual points. 
The 6-parametric invariance of the 
Kepler dynamics  is - up
to the trivial solution  with position and  momentum  
$(\rvec x,\rvec p)=(0,0)$  - broken to
a 1-parametric fixgroup
symmetry
since the choice of
angular momentum and Lenz-Runge vectors $(\rvec\cl L,\rvec\cl P)$ (six parameters) with 
$\rvec\cl L\rvec\cl P=0$ (one condition) to determine an orbit comes 
from a 5-parametric manifold
\begin{eq}{rlll}
H&=\SO(2),&\SO_0(1,1),&\R\cr 
G/H&=\SO(4)/\SO(2),&\SO_0(1,3)/\SO_0(1,1),&[\SO(3)\sx\R^3]/\R\cr
\end{eq}The nontrivial
orbits $t\mape\rvec x(t)$  in position
as quotient groups $\R/K$ of time  are isomorphic to
the characterizing  fixgroups $H$  in the Kepler groups.
In all three cases, the solution degeneracy is  
$G/H\cong\SO(3)\x\Om^2$ with
all three compact rotation parameters  for angular momentum  $\rvec{\cl L}$
and two compact parameters 
for the direction of the Lenz-Runge vector $\rvec{\cl P}$.

\subsection{Two Sided Contraction to the Free Theory}

The  complex 6-di\-men\-sional group $\SO(\C^4)$,
 e.g. in the 
 defining 4-di\-men\-sional  re\-pre\-sen\-ta\-tion for its Lie algebra 
\begin{eq}{l}
l(\rvec\phi,\rvec\psi)=\rvec\phi\rvec L+\rvec\psi\rvec B={\scriptsize\left(\begin{array}{c|ccc}
0&\psi_1&\psi_2&\psi_3\cr\hline
\psi_1&0&\phi_3&-\phi_2\cr
\psi_2&-\phi_3&0&\phi_1\cr
\psi_3&\phi_2&-\phi_1&0\cr
\end{array}\right)
}\in\log\SO(\C^4)
\end{eq}with  complex rank 2 and the corresponding two invariant bilinear forms $\ka_{1,2}$ 
from the coefficients of the characteristic polynomial
\begin{eq}{rl}
\det[l(\rvec\phi,\rvec\psi)-\la\bl1_4]
&=\la^4+\la^2(\rvec\phi^2-\rvec\psi^2)-(\rvec\phi\rvec\psi)^2\cr
\then& \left\{\begin{array}{rl}
\ka_1(l,l)&=-\tr l\o l=\rvec\phi^2-\rvec\psi^2\cr
\ka_2(l,l)&=\sqrt{-\det l}=\rvec\phi\rvec\psi\end{array}\right.
\end{eq}has as  real forms the
 groups
 \begin{eq}{ll}
 \SO_0(1,3)&\hbox{with }\phi_a\in\R,~\psi_a\in\R\cr
 \SO(4)&\hbox{with }\phi_a\in\R,~ \psi_a=i\chi_a\in i\R
 \end{eq}Both groups are expansions of
the non-semisimple Euclidean group $\SO(3)\sx\R^3$
\begin{eq}{l}
{\scriptsize\left(\begin{array}{c|ccc}
0&0&0&0\cr\hline
\xi_1&0&\phi_3&-\phi_2\cr
\xi_2&-\phi_3&0&\phi_1\cr
\xi_3&\phi_2&-\phi_1&0\cr
\end{array}\right)
}\in\log[\SO(3)\sx\R^3]
\end{eq}with the  
`boosts' $\rvec\psi$  or  the  additional `internal rotations' 
$\rvec\chi$ as `unflattened, expanded 
translations'.  

Vice versa, the translations $\R^3$  arise by  
In\"on\"u-Wigner contraction\cite{INWIG} 
to  the Galilei group
as tangent space both  of the  compact
3-sphere   $\Om^3\cong\SO(4)/\SO(3)\cong\SO(2)\x\Om^2$,
related to bound structures,  and of the  noncompact hyberboloid 
$\cl Y^3\cong\SO_0(1,3)/\SO(3)\cong\SO_0(1,1)\x\Om^2$,
related to scattering
structures,
to the free theory  with  
non-semisimple symmetry
\begin{eq}{ccccc}
\hbox{spherical}&&\hbox{flat}&&\hbox{hyperbolic}\cr
{\scriptsize\pmatrix{
\SO(4)\cr\cup\cr\SO(4)/\SO(3)\cr\cup\cr\SO(2)\cr}}
&\stackrel{\eta \to 0}\longrightarrow&
{\scriptsize\pmatrix{
\SO(3)\sx\R^3\cr\cup\cr\R^3\cr\cup\cr\R\cr}}
&\stackrel{\eta \to0}\longleftarrow&
{\scriptsize\pmatrix{
\SO_0(1,3)\cr\cup\cr\SO_0(1,3)/\SO(3)\cr\cup\cr\SO_0(1,1)\cr}}
\end{eq}

The contraction procedure will be given explicitly for the decisive abelian
subgroups (last line):
The relevant contraction parameter $\eta^2={b^2\over a^2}= 2|E|L^2$ 
with energy and angular momentum 
is the ratio of the units
for the two directions
in the conic sections. It is the analogue to the ratio of a time 
unit to a position unit
${1\over c^2}={\tau^2\over \ell^2}$ in the archetypical Wigner-In\"on\"u
contraction. 
For the Coulomb interaction
$V(r)={1\over 4\pi\ep_0}{z_1z_2e^2\over r}$
with $\de=\ep(z_1z_2)=\pm 1$ 
and  the unit for the product of energy and angular momentum
 $[EL^2]={me^2\over\ep_0}$
the  contraction limit is realizable by ${e^2\over \ep_0}\to 0$.
For the planetary system with gravitational interaction $-{G_NMm\over r}$
and  $[EL^2]=m(G_NMm)^2$
the  contraction limit is realized by $G_N\to 0$.

The ratio of the units is used
for a renormalization of the Lie parameters
$(\chi,\psi)\stackrel{\eta}\mape\xi$ 
- the analogue for the reparametrization from rapidity
to velocity $\tanh\psi={v\over c}$. 
The contraction of the length ratio  
$\eta={b\over a}=L\sqrt{2|E|}\to 0$ is the analogue to 
the contraction to an infinite velocity ${1\over c}={\tau\over \ell}\to 0$
\begin{eq}{rrl}  
\left.\begin{array}{r}
E<0\cr
\tan\chi=\eta \xi\cr
\end{array}\right\}
:&\hskip-2mm\SO(2)&\hskip-2mm\ni{\scriptsize\pmatrix{
\cos\chi&\eta ~i\sin\chi\cr
{1\over \eta }~i\sin\chi&\cos\chi\cr}}\cr

&&={1\over \sqrt{1+\eta^2\xi^2 }} {\scriptsize\pmatrix{
1&\eta^2i\xi\cr
i\xi&1\cr}}
\to{\scriptsize\pmatrix{
1&0\cr
i\xi&1\cr}}\hbox{ for }\eta\to 0\cr
\left.\begin{array}{r}
E>0\cr
\tanh\psi=\eta \xi\cr
\end{array}\right\}
:&\hskip-2mm\SO_0(1,1)&\hskip-2mm\ni
{\scriptsize\pmatrix{
\cosh\psi&\eta ~\sinh\psi\cr
{1\over\eta }~\sinh\psi&\cosh\psi\cr}}\cr
&&={1\over \sqrt{1-\eta^2\xi^2 }} {\scriptsize\pmatrix{
1&\eta^2\xi \cr
\xi&1\cr}}\to {\scriptsize\pmatrix{
1&0\cr
\xi&1\cr}}\hbox{ for }\eta\to 0\cr

\end{eq}The contracted additive group $\R$
comes in the multiplicative re\-pre\-sen\-ta\-tion 
with
nilpotent operations, 
typical for the non-semisimplicity of semi-direct groups
\begin{eq}{l}
\R\ni \xi\mape {\scriptsize\pmatrix{1&0\cr \xi&1\cr}}=\exp
{\scriptsize\pmatrix{0&0\cr \xi&0\cr}}
\hbox{ with }{\scriptsize\pmatrix{0&0\cr \xi&0\cr}}^2=0
\end{eq}The contraction limit  describes the line orbits of the free theory,
not the parabolas of the Kepler potential.

\section{Quantum Kepler Bound States}

In the quantum case,
the Kepler Hamiltonian, angular momentum  and  Lenz-Runge vector 
\begin{eq}{ll}
H={\rvec p ^2\over2}+{\de \over r },~~
\rvec {\cl L}=\rvec x\x\rvec p,~~
\rvec{\cl  P}=
{\rvec p\x\rvec {\cl L}-\rvec {\cl L}\x\rvec p\over2}
+\de {\rvec x\over r }\cr
\end{eq}give  the same three Lie algebra
structures as in the classical case
\begin{eq}{l}
\com H{\rvec {\cl L}}=0,~~\com H{\rvec {\cl P}}=0,~~
\left\{\begin{array}{rl}
\com{{i\cl L}^a}{{i\cl L}^b}&=-\ep^{abc}{i\cl L}^c\cr
\com{{i\cl L}^a}{{i\cl P}^b}&=-\ep^{abc}{i\cl P}^c\cr
\com{{i\cl P}^a}{{i\cl P}^b}&=2H\ep^{abc}{i\cl P}^c\cr
\end{array}\right.
\end{eq}The additional $i$-factor is related to the different dual
normalization in the Poisson bracket $[p,x]_P=1$ and the quantum commutator
$i[p,x]=1$.

Again, the squares of the angular momentum and the Lenz-Runge vector
determine the Hamiltonian
\begin{eq}{c}
\rvec {\cl P}^2=1+2H(\rvec {\cl L}^2+1)\then
-{1\over 2H}=1+\cl L^2-{\cl P^2\over2H}
\end{eq}with an additional constant compared to the classical case.

For   $\de=-1$ and negative energies  
one has re\-pre\-sen\-ta\-tions of the
compact symmetry Lie algebra 
\begin{eq}{l}
\spec H\ni E<0:\left\{\begin{array}{l}
\rvec {\cl B}={\rvec {\cl P}\over\sqrt{-2H}},~~
\rvec {\cl J}_\pm={\rvec {\cl L}\pm\rvec {\cl B}\over2}\cr
\com{{i\cl J}^a_\pm}{{i\cl J}^b_\pm}
=-\ep^{abc}{i\cl J}^c_\pm,~~\com{{\cl J}^a_+}{{\cl J}^b_-}=0\cr
\hbox{invariants: }\rvec {\cl L}^2+\rvec {\cl B}^2=
2(\rvec {\cl J}_+^2+\rvec {\cl J}_-^2),~~
\rvec {\cl L}\rvec {\cl B}=
\rvec {\cl J}_+^2-\rvec {\cl J}_-^2\cr
\end{array}\right.\cr\log[\SO(3)\x\SO(3)]\cong A_1^c\pl A_1^c\cr
 \end{eq}

The weight diagrams of the irreducible $\SU(2)\x \SU(2)$-re\-pre\-sen\-ta\-tions
\begin{eq}{rl}
\irrep[\SU(2)\x\SU(2)]&=\irrep\SU(2)\x \irrep\SU(2)\cr
&=\{(2J_1,2J_2)
\mid J_{1,2}=0,{1\over2},1,\dots\}
\end{eq}occupy $(1+2J_1)(1+2J_2)$
 points of a  rectangular grid.

The two invariants determine the occurring re\-pre\-sen\-ta\-tions.
The triviality of the  invariant  $\rvec{\cl L}\rvec{\cl P}=0$
(classical orthogonality 
of angular momentum and Lenz-Runge vector)
`synchronizes'  the centers $\I_2=\{\pm1\}$ 
of both  $\SU(2)$'s
(central correlation - two cycles give one  bicycle)  and leads to the relevant group $\SO(4)$
\begin{eq}{l}
{\SU(2)\x\SU(2)\over \I(2)}\cong\SO(4)~~~
\hbox{with }\I(2)=\{(1,1),(-1,-1)\}\subnoteq\SU(2)\x \SU(2)
\end{eq}It
enforces even the equality of both $\SU(2)$-invariants  $J_+=J_-=J$ 
\begin{eq}{l}
0=\rvec {\cl L}\rvec {\cl B}=
\rvec {\cl J}_+^2-\rvec {\cl J}_-^2
\then \angle{ \rvec {\cl J}_+^2}=\angle{\rvec {\cl J}_-^2}=J(1+J)
,~~
J=0,{1\over2},1,{3\over2},\dots\cr
\end{eq}Therefore 
 the energy degenerated  re\-pre\-sen\-ta\-tions are of the
  type $(2J,2J)$,  
  the multiplets 
of both $A_1^c$-re\-pre\-sen\-ta\-tions
have equal
di\-men\-sio\-n $1+2J$. The $\SU(2)$-multiplet dimension 
is  the {principal quantum number}  $k=1+2J$.
The weight diagrams  occupy $(1+2J)^2$ points of a square   grid
\begin{eq}{l}
\irrep\SO(4)\Big|_{\rm{Kepler}}=\{(2J,2J)\mid J=0,{1\over2},1,\dots\},~~
(2J,2J)={\OD^{2J}}(1,1)
\end{eq}The Kepler re\-pre\-sen\-ta\-tions are the totally symmetrized
products of the  
defining $\SO(4)$-re\-pre\-sen\-ta\-tion 
$(1,1)$.

The  energy eigenvalues
are given with the value
 of the Casimir operator
\begin{eq}{rl}
-{1\over 2\angle H}&=1+2\angle{\rvec {\cl J}_+^2+\rvec {\cl J}_-^2}=
1+4J(1+J)
,~~
J=0,{1\over2},1,{3\over2},\dots\cr
 E_k&=-{1\over 2 k^2},~~
\hbox{ multiplicity: }k^2=(1+2J)^2=1,4,9,16,\dots

\end{eq}

The $\SO(4)$-re\-pre\-sen\-ta\-tions are decomposable
with respect to the position  rotation $\SO(3)$-properties 
into irreducible re\-pre\-sen\-ta\-tions of dimension
$(1+2L)$ with  integer $L=0,1,\dots$ for 
$\rvec\cl L=\rvec\cl J_++\rvec\cl J_-$
\begin{eq}{rl}
(2J,2J)&\stackrel{\SO(3)}\cong{\PL_{L=0}^{2J}}[2L]\cr
2J=L+N&\then(L,N)=(2J,0),(2J-1,1),\dots,(0,2J)\cr
\end{eq}The Lenz-Runge invariance related difference
$2J-L=N$ characterizing the classes  $\Om^3\cong \SO(4)/\SO(3)$ 
is  the {radial quantum number or knot number}.

\section
[Orbits of 1-Dimensional Position]
{Orbits of 1-Dimensional Position}

In contrast to the classical framework, the time orbits in quantum theory are not 
 valued in position space.
For bound states they are valued in a Hilbert space. 
In addition to the time orbits
there are  - in a Schr\"o\-din\-ger picture - 
position orbits effected by complex valued 
position re\-pre\-sen\-ta\-tions  $\psi$ -  
for bound states\footnote{\scriptsize
The function space $L^p_{d\mu}(S,\C)$ will be denoted by $L^p(S)$
if the positive $S$-measure $d\mu$ is unique up to a factor, e.g. for a locally
compact group with Haar measure.} 
\begin{eq}{l}
(t,x)\mape \Psi(t,x)=e^{iEt}\psi(x)\in\C,~~e^{iEt}\in\U(1),~~\psi\in L^2(\R^s)
\end{eq}

In the energy eigenvalue  problem 
 of a  Hamiltonian $H={p^2\over 2}+V$, parametrized with 1-di\-men\-sional 
position translations 
\begin{eq}{l} 
ip\cong{d\over dx}=d_x:~~[-{1\over2}d_x^2+V(x)]\psi(x)=E \psi(x) 
\end{eq}the real energies for the time translation eigenvalues
are given in terms of { real or imaginary `momenta'} for the position
translation eigenvalues.
The differential equation for constant potential 
with position translation invariance of the Hamiltonian
\begin{eq}{l} 
H_0={p^2\over 2}+V_0,~~[H_0,p]=0,~~

 [d_x^2+2(E-V_0)] \psi_0(x)=0
\end{eq}is solved by two types of
the  re\-pre\-sen\-ta\-tion matrix elements 
  of the noncompact
position group $\R$ - either
with imaginary or with real eigenvalues, representing quotient groups of
position $\R$, of the same type as for the classical time orbits
in position above  
 \begin{eq}{l}
\R \ni x\mape  \left\{\begin{array}{rl}
{\scriptsize\pmatrix{\cos qx&i\sin qx\cr i\sin qx&\cos qx\cr}}
&\cong {\scriptsize\pmatrix{e^{iqx}&0\cr0&e^{-iqx}\cr}}
\in\SO(2)\subnoteq \SU(2)\cr
E-V_0&={q^2\over2}>0\hbox{ (free scattering waves)}\cr
{\scriptsize\pmatrix{
\cosh Qx&\sinh Qx\cr\sinh Qx&\cosh Qx\cr}}
&\cong {\scriptsize\pmatrix{
e^{Qx}&0\cr0&e^{-Qx}\cr}}
\in\SO_0(1,1)\subnoteq \SU(1,1)\cr
E-V_0&=-{Q^2\over2}<0\hbox{ (bound waves)}\cr
\end{array}\right.
\end{eq}Matrix elements of reducible 
nondecomposable representations come with
strict\-ly positive position translation  powers (nildimensions) 
$x^Ne^{\pm iqx}$ and $x^N e^{\pm Qx}$,
$N=1,2,\dots$.
These representations are indefinite unitary.
The order structure of the reals, i.e. the bicone property
$\R=\R_+\uplus\R_-$, can be represented with additional
factors $\vth(\pm x)$ and $\ep(x)$, e.g. in  $|x|=\ep(x)x$.

The groups realized by time $\R$ and position $\R$ orbits
are the product of $ \U(1)$ for time
with the  corresponding 
real 1-dimensional  groups for position
\begin{eq}{rrl}
H_0={p^2\over 2}+V_0:&
\R\x\R&\map \U(1)\x {\scriptsize\pmatrix{
\SO(2)\cr
\SO_0(1,1)\cr
}}\cr
{\scriptsize\pmatrix{
\hbox{free waves}\cr
\hbox{bound waves}\cr
}}:&
(t,x)&\mape e^{iEt} {\scriptsize\pmatrix{ e^{\mp iqx}\cr e^{-|Qx|}\cr}},
~~ E-V_0={\scriptsize\pmatrix{{\rvec q^2\over 2}\cr-{Q^2\over 2}\cr}}
\end{eq}The re\-pre\-sen\-ta\-tion invariants for time
(energy) and position are related to each other.
The difference $E-V_0$
connects with each other the
eigenvalues $iE$ for the time translation representation
with the energy $E$ and the eigenvalues $iq$ for the 
position translation representation with the
momentum $q$. It is the nonrelativistic precursor of  
the relativistic energy-momentum relation  
$\rvec q^2=q_0^2-m^2$ as used for quantum fields.
 The
{compact position representations} in $\SO(2)$ 
for{ scattering waves}  come for positive kinetic energies, i.e.
for
$E>V_0$ ($q_0^2>m^2$)
with real momentum  $q^2>0$ 
(`on shell' real particles) and imaginary eigenvalue $i q$. 
The{ noncompact position   representations }in 
$\SO_0(1,1)$ for{ bound waves} 
come for $E<V_0$ ($q_0^2<m^2$) with imaginary `momentum' $(iQ)^2<0$ 
(`off shell' virtual particles)
 and real eigenvalue $|Q|$.

The conjugation and, therewith, the scalar product (probability amplitude)
for  Hilbert spaces  is determined by the
positive  unitary group  $\U(1)\ni e^{iEt}$ with the
time re\-pre\-sen\-ta\-tions, not by the group with the position 
translation 
re\-pre\-sen\-ta\-tions which for $\U(1)$-time re\-pre\-sen\-ta\-tions are
definite unitary $\SU(2)$ for free scattering waves and indefinite unitary 
$\SU(1,1)$ for  bound waves.

Bound waves will be 
defined as square integrable functions, i.e.
as elements of the position function Hilbert space.
Free scattering waves are no position Hilbert space vectors, they are
tempered distributions
\begin{eq}{l}
\{x\mape e^{-|Qx|}\}\in L^2(\R),~~
\{x\mape  e^{-iqx}\}\in\cl S'(\R) \cr
\end{eq}Hilbert space vectors for scattering states need
square integrable momentum wave packets $f$
in the Fourier isomorphism
\begin{eq}{l}
L^2_{dp}(\R)\ni f\lrmap \psi\in L^2_{dx}(\R),~~\psi(x)=\int {dp\over2\pi}~ f(p)e^{ipx}
\end{eq}Integrals without boundary go over the full integration space,
here $\int dq=\int_{-\infty}^\infty dq$.
Both the free scattering and the  bound  waves with  compact and 
noncompact 
re\-pre\-sen\-ta\-tion matrix elements  
can be  Fourier expanded with the unitary position characters $e^{ipx}\in\U(1)$
\begin{eq}{rl}
e^{-|Qx|}&=\int
{dp\over \pi}~{|Q|\over  p^2+Q^2}e^{-ipx}\cr
e^{-iqx}&=\int dp~\de(p-q)e^{-ipx}
=\oint {dp\over 2i\pi}{1\over p- q}e^{-ipx}
\cr

\end{eq}The irreducible exponentials  
come as residues 
(counterclockwise integration $\oint$) of a real momentum pole $p= q$  for the compact re\-pre\-sen\-ta\-tions 
and an  imaginary `momentum' pole $p=-i|Q|$
for the noncompact ones.

\section
[Orbits of 3-Dimensional Position]
{Orbits of 3-Dimensional Position}

The 3-di\-men\-sional position translations with  
rotation group $\SO(3)$ action are,  in polar coordinates, 
the product $\R^3\cong \Om^2\x \R_+$ of the totally ordered cone $\R_+$
with the radial translations  
and the compact $2$-sphere.
Both factors will be presented by corresponding orbits.
The 1-dimensional case is embedded as abelian substructure
\begin{eq}{l}
\R\sub\SO(s)\sx\R^s,~~\R^s\cong\Om^{s-1}\x\R_+,~~s\ge 1 
 \end{eq}The $0$-sphere consists of two points 
 $\Om^0=\{\pm1\}\cong\I(2)$ - forwards and backwards.

In a  derivative re\-pre\-sen\-ta\-tion,
acting upon differentiable complex functions 
the radial and angular momentum squares look
as follows
\begin{eq}{rl}
\rvec p^2&=
p_r^2+{\rvec {\cl L}^2\over r^2}
,~~[\rvec p^2,\rvec{\cl L}]=0\cr
\rvec{\cl L}^2&\cong
{1\over\sin^2\th}[(\sin\th{\p\over\p\th})^2
+({\p\over\p\phi})^2]=
{\p^2\over\p\th^2}+{1\over\tan\th}{\p\over\p\th}
+({1\over\sin\th}{\p\over\p\phi})^2
\end{eq}

The
{ harmonic polynomials}  
as product of spherical harmonics with
the corresponding  radial power
 are homogeneous in the coordinates $\rvec x$. They are defined
 also for $\rvec x\to0$ in contrast to the spherical harmonics 
 $\ro Y^L_m$  with the 
 $r\to 0$   ambiguity in ${\rvec x\over r}$.
 The harmonic polynomials are  eigenfunctions  
for translation invariant $\rvec p^2$ with  trivial value $\rvec q^2=0$
\begin{eq}{l}
(\rvec x)^L_m=r^L\ro Y^L_m(\phi,\th):~~~\left.\begin{array}{rl}
\rvec{\cl L}^2\ro Y^L_m(\phi,\th)
&=L(1+L)\ro Y^L_m(\phi,\th)\cr
p_r^2r^L&=-L(1+L)r^{L-2}\cr
\end{array}\right\}\then \rvec \p^2 (\rvec x)^L_m=0\cr
\end{eq}For fixed $L$ and any $r$
they span a space acted upon with
an  irreducible  rotation group
 $\SO(3)$-representation. 

A rotation invariant Hamiltonian is decomposable 
into the generators $H_L$  for each angular momentum
\begin{eq}{l}
H={\PL_{L=0}^\infty}H_L\hbox{ with }\left\{\begin{array}{rl}
H&={\rvec p^2\over 2}
+V(r),~~[H,\rvec{\cl L}]=0\cr
H_L&={p_r^2\over 2}+{L(1+L)\over 2r^2}
+V(r)\cr
&=
r^n[-{1\over 2}d_r^2-{1+n\over r}d_r
+{L(1+L)-n(1+n)\over 2r^2}+V(r)]{1\over r^n}\cr
n&=0,\pm 1,\pm 2,\dots\cr
\end{array}\right.
\end{eq}Attention has to be paid to the 
small distance $r\to 0$ behavior, prepared  with
the powers $r^n$ and used below.

The radial functions $\{\psi_L^m\}$ 
are radial translation re\-pre\-sen\-ta\-tion
matrix elements. The noncompact  re\-pre\-sen\-ta\-tions
for bound waves act on a Hilbert space, the 
compact ones  are  scattering distributions. The
irreducible  ones  are given by   
\begin{eq}{l}
\R_+\ni r\mape \left\{\begin{array}{lll}
e^{-|  Q |r}&\hbox{with }(d_r^2 -  Q ^2)e^{-|  Q |r}&=0 \cr
e^{\pm i|\rvec q|r}&\hbox{with }(d_r^2 +\rvec q^2)e^{\pm i|\rvec q|r}&=0\cr
\end{array}\right.
\end{eq}Their Fourier transforms  involve a dipole at imaginary `momenta'
$|\rvec p|=\pm i|  Q |$ for
bound states (noncompact, hyperbolic) and a pole
for free scattering states (compact, spherical) 
\begin{eq}{rll}
e^{-|  Q |r}&=\int{d^3p\over \pi^2}\hfill
{|  Q |\over (\rvec p^2+  Q ^2)^2}e^{-i\rvec p\rvec x}
&\hbox{from } L^2(\R^3)\cr
{\sin|\rvec q|r\over  |\rvec q| r}
=\int {d^3p\over 2\pi |\rvec q| }
~\de(\rvec p^2-\rvec q^2)e^{-i\rvec p\rvec x}
&=\oint {d^3p\over 4i\pi^2 |\rvec q| }
~{1\over \rvec p^2-\rvec q^2}e^{-i\rvec p\rvec x}
&\hbox{from } \cl S'(\R^3)\cr
\end{eq}With the integral around all poles
\begin{eq}{l}
a\in\R:~~ \de(a)\to\oint{1\over 2i\pi}{1\over a}\cr
\end{eq}the re\-pre\-sen\-ta\-tions can be written as residues
at the negative and positive invariants.

\subsection{Scattering Distributions}

Compact radial re\-pre\-sen\-ta\-tions
are spread to the 2-sphere with a  factor ${1\over r}$
and the  separation of 
the spherical harmonics
\begin{eq}{rl}
 \psi(\rvec x)
={\SUM_{L=0}^\infty}~{\SUM_{m=-L}^L}
 ({\rvec x\over r})^L_m {D_L(r)\over r}
&\then
[d_r^2-{L(1+L)\over r^2} +2(E-V(r))]D_L(r)=0\cr

D_L(r)=r\psi_L(r)&
\end{eq}

An irreducible compact radial position re\-pre\-sen\-ta\-tion
requires a  constant potential
\begin{eq}{l}
D_0(r)=e^{\pm i|\rvec q|r}\then H={\rvec p^2\over2}+V_0,~~E-V_0= {\rvec q^2\over 2}\cr
\end{eq}The solutions for general $L$ use the spherical Bessel functions
\begin{eq}{l}
[d_r^2-{L(1+L)\over r^2}+\rvec q^2]D_L(r)=0\then D_L(r)=|\rvec q|rj_L(|\rvec q|r)
\end{eq}which have to be considered as tempered distributions, e.g.
\begin{eq}{l}
{\scriptsize\pmatrix{
j_0(r)\cr
{\rvec x\over r}j_1(r)\cr}}
={\scriptsize\pmatrix{
{\sin r\over r}\cr
{\rvec x\over r}{\sin r-r\cos r \over r^2}\cr}}
=\int{d^3p\over 2\pi}
{\scriptsize\pmatrix{
1\cr i\rvec p\cr}}\de(\rvec p^2-1)e^{-i\rvec p\rvec x}
\end{eq}They act via Fourier-Bessel transforms
$\int_0^\infty dq f_L(q) q j_L(qr)$ upon momentum wave packets 
$f_L\in L^2(\R_+)$.

\subsection{Multipole Energy Measures for the Kepler Potential}

Noncompact radial representations come
with the  separation of  the  harmonic 
polynomials for the irreducible $\SO(3)$-representations. This leads to
the Schr\"o\-din\-ger equations
for the position re\-pre\-sen\-ta\-tion
matrix elements 
 \begin{eq}{rl}
 \psi(\rvec x)
 ={\SUM_{L=0}^\infty}~{\SUM_{m=-L}^L}
 (\rvec x)^L_m d_L(r)&\then
[d_r^2+{2(1+L)\over r}d_r
+2(E-V(r))]d_L(r)=0\cr
r=|\rvec x|&
\end{eq}

For an  irreducible noncompact radial position re\-pre\-sen\-ta\-tion
as bound solution
 an attractive  Kepler
potential is necessary with  an associate angular momentum dependent
momentum unit - a `ground state' for each angular momentum
\begin{eq}{l}
d_L(r)=e^{-|  Q |r}\then V(r)=-{1\over r}
\then (1+L)Q_L=(1+L)\sqrt{-2 E_{L0}}=1
\end{eq}There arises the 
quantum analogue to the  classical
parameter $L\sqrt{-2E}$.
For the attractive Kepler potential and negative energy
the imaginary radial `momentum' $\sqrt{2E}$ is `quantized' (integer wave
numbers $k$)
in  the radial bound  waves
\begin{eq}{l}
\begin{array}{r}[d_r^2+{2(1+L)\over r}d_r
+2(E+{1\over r})]d_L(r)=0\cr
{\rho(r)\over2}=|Q|r\cr
E=-{Q^2\over2}<0\cr
\end{array}\then\left\{\begin{array}{rl}
d_L(r)&=\ro L^N_{1+2L}(\rho)e^{-{\rho\over2}}\cr
{1\over Q_k}=k&=1+2J=1+L+N\cr
&\hfill L,N=0,1,\dots\cr
\psi_{Lm}^{2J}(\rvec x)&\sim
 ({\rvec x\over k})^L_m~
 \ro L_{1+2L}^N({2r\over k})~e^{-{r\over k}}
\cr
\end{array}\right.\cr

\end{eq}

The $\SO(4)$ multiplets
comprise all wave functions $\psi^{2J}$ with 
equal sum $L+N=2J$ for the principal quantum number
$k=1+2J$ with  angular momentum $(1+2L)$-multiplets for $\SO(3)$
and  radial quantum  numbers $N$ for Lenz-Runge classes 
$\Om^3\cong\SO(4)/\SO(3)$.
The products of harmonic and Laguerre polynomials
are related to the $\Om^3$-analogue of the spherical harmonics on $\Om^2$.
The Fourier transformations
of the bound state wave functions
involve the $\SO(3)$-invariant momentum measure 
${d^3p\over 4\pi^2}({2\over 1+\rvec p^2})^2$ 
with dipoles at the
negative invariants (imaginary `momentum' eigenvalues)
\begin{eq}{l}
 Q=1:~~
\psi_0^0
(\rvec x)\sim
e^{-Qr}
=\int {d^3p\over Q^34\pi^2}
({2 Q^2\over Q^2+\rvec p^2})^2e^{-i\rvec p\rvec x}
=\int {d^3p\over 4\pi^2}
({2 \over 1+\rvec p^2})^2e^{-i\rvec pQ\rvec x}
\end{eq}The $k=2$ bound states quartet
 have tripole vector measure
\begin{eq}{rl}
Q={1\over 2}:~~{\scriptsize\pmatrix{
\psi_0^1\cr\psi_1^1\cr}}(\rvec x)\sim
{\scriptsize\pmatrix{
\hfill{1\over 4}\ro L^1_1(2Qr)\cr
\hfill {Q\rvec x\over2}~\ro L^0_2(2Qr)\cr}}
&=
{\scriptsize\pmatrix{
{1-Qr\over2}\cr
{Q\rvec x\over2}\cr}}e^{-Qr}\cr
&=\int {d^3p\over4\pi^2}
({2\over1+\rvec p^2})^3e^{-i\rvec pQ\rvec x}
{\scriptsize\pmatrix{
{\rvec p^2-1\over 2}\cr
i\rvec p\cr}}\cr
\end{eq}The normalized 4-vector on the 3-sphere under the integral 
\begin{eq}{rl}
{2\over1+\rvec p^2}
{\scriptsize\pmatrix{
{\rvec p^2-1\over 2}\cr
i\rvec p\cr}}=
{\scriptsize\pmatrix{q_0\cr i\rvec q\cr}}=
{\scriptsize\pmatrix{\cos\chi \cr i\sin\chi{\rvec p\over |\rvec p|}
\cr}}
\in\Om^3
\hbox{ with }q^2=q_0^2+\rvec q^2=1
\cr
\end{eq}is the analogue 
$\ro Y^{(1,1)}(q)\sim
{\scriptsize\pmatrix{q_0\cr i\rvec q\cr}}\in\Om^3$
to the normalized  3-vector 
$\ro Y^1({\rvec p\over |\rvec p|})\sim {\rvec p\over |\rvec p|}\in\Om^2$ 
which builds the  2-sphere harmonics 
$\ro Y^L({\rvec p\over |\rvec p|})\sim
({\rvec p\over |\rvec p|})^L$. It
depends on three angles for  $\Om^3$ which can be
parametrized with a 3-vector $\rvec p\in\R^3$. 
The higher order $\Om^3$-harmonics
arise from the totally symmetric traceless
products 
$\ro Y^{(2J,2J)}(q)\sim(q)^{2J}$, e.g.
 the nine independent components  in the 
 $(4\x4)$-matrix 
\begin{eq}{rl}
\ro Y^{(2,2)}(q)&
\sim (q)^2_{jk}=q_jq_k-{\de_{jk}\over 4}
\cong {\scriptsize\left(\begin{array}{c|c}
{3q_0^2-\rvec q^2\over 4}&iq_0q_a\cr\hline
iq_0q_b&
q_aq_b-{\de_{ab}\over 4}\cr
\end{array}\right)}\cr
q_aq_b-{\de_{ab}\over 4}
&=q_aq_b-{\de_{ab}\over 3}\rvec q^2-{\de_{ab}\over 3}{3q_0^2-\rvec q^2\over4}
\hbox{ with }q^2=1\cr
\end{eq}which are used for
the $k=3$ bound states nonet
 with quadrupole tensor measure
\begin{eq}{rl}
Q={1\over 3}:~~{\scriptsize\pmatrix{
\psi_0^2\cr
\psi_1^2\cr
\psi_2^2\cr}}(\rvec x)&\sim
{\scriptsize\pmatrix{
\hfill{1\over 3}\ro L^2_1(2Qr)\cr
\hfill {Q\rvec x\over2}~{1\over 6}\ro L^1_3(2Qr)\cr
{Q^2\over2}(\bl 1_3{r^2\over 3}-\rvec x\ox\rvec x)\ro L^0_5(2Qr)\cr}}
=
{\scriptsize\pmatrix{
1-2Qr+{2Q^2r^2\over 3}\cr
{2-Qr\over 3}{Q\rvec x\over2}\cr
{Q^2\over2}(\bl 1_3{r^2\over 3}-\rvec x\ox\rvec x)
}}e^{-Qr}\cr
&=\int {d^3p\over 4\pi^2}
({2 \over 1+\rvec p^2})^4e^{-i\rvec pQ\rvec x}
{\scriptsize\pmatrix{
3({\rvec p^2-1\over 2 })^2-\rvec p^2\cr
i\rvec p~{\rvec p^2-1\over 2 }\cr
3\rvec p\ox\rvec p-\bl 1_3\rvec p^2
\cr}}\cr
&=\int {d^3p\over 4\pi^2}
({2 \over 1+\rvec p^2})^2e^{-i\rvec pQ\rvec x}
{\scriptsize\pmatrix{
3q_0^2-\rvec q^2\cr
 iq_0\rvec q\cr
3\rvec  q\ox\rvec q-\bl 1_3\rvec q^2 \cr}}\cr
\end{eq}and in general
with $2J$-dependent multipole measure of the momenta
\begin{eq}{l}
Q={1\over 1+2J},~J=L+N:~~
\psi_L^{2J}(\rvec x)
\sim
\int {d^3p\over4\pi^2}
({2\over 1+\rvec p^2})^2e^{-i\rvec pQ\rvec x}
(q)^{2J},~~ q=
{2\over1+\rvec p^2}
{\scriptsize\pmatrix{
{\rvec p^2-1\over 2}\cr
i\rvec p\cr}}
\end{eq}

\subsection{Nonrelativistic Color Symmetry}

Separating the harmonic polynomials with a squared radial 
dependence, there arise the  radial equations
 \begin{eq}{rl}
 \psi(\rvec x)
 ={\SUM_{L=0}^\infty}~{\SUM_{m=-L}^L}
 (\rvec x)^L_m \De_L(\rho)
 &\then[\rho d_\rho^2
+({3\over2}+L-\rho)d_\rho+{E-V(r)\over2}]\De_L(\rho)=0\cr
\rho=r^2=\rvec x^2&
\end{eq}

An irreducible exponential with squared radial dependence 
as solution determines the harmonic oscillator potential, normalized\footnote{\scriptsize
Usually, the additive term $V_0={3\over2} Q ^2$ is introduced as
ground state  energy.}
 with a  momentum unit $|Q|$
\begin{eq}{l}
\De_L(r^2)=e^{-{ Q ^2r^2\over 2}}\then V(r)= {(Q_Lr)^2\over2}Q_L^2,~~
E=({3\over2}+L) Q_L^2
\end{eq}The momentum unit can be chosen
$L$-independent, e.g.  $Q_L=1$.
The general solutions come with Laguerre polynomials
of degree $N$ and - in contrast to the Kepler bound state solutions
with principal quantum number dependent exponentials $e^{-{r\over k}}$ -
with one exponential only 
\begin{eq}{l}
[d_r^2+{2(1+L)\over r}d_r+2(E- {r^2\over2})]\De_L(r^2)=0

\then\left\{\begin{array}{rl}
\De_L(r^2)&=L^N_{{1+2L\over2}}( r^2)e^{-{ r^2\over2}}\cr
E_k&={3\over2}+k ={3\over2}+L+2N \cr
&\hfill L,N=0,1,\dots\cr
\psi_{Lm}^{k}(\rvec x)&\sim
(\rvec x)^L_m~\ro L^N_{{1+2L\over2}}(r^2)~e^{-{r^2\over2}}
\end{array}\right.
\end{eq}

The  harmonic oscillator Hamiltonian
\begin{eq}{l}
H={\rvec p^2+ \rvec x^2\over 2}= 
{\acom{\ro u^a}{\ro u_a^\star}\over 2}\hbox{ with }

\hbox{creation operators }\ro u^a={  x^a-ip^a\over\sqrt2},~~
[\ro u_b^\star,\ro u^a]=\de_a^b\cr
\end{eq}generates time orbits of the Hilbert vectors with $k$ quanta. They 
have a Schr\"o\-din\-ger representation as position orbits 
\begin{eq}{rl}
\rstate {a_1,\dots a_k}&={\ro u^{a_1}\cdots\ro u^{a_k}\over \sqrt{k!}}\rstate
0\cong\{\rvec x\mape  x^{a_1}\cdots x^{a_k}e^{-{r^2\over2}}\}\in
L^2(\R^3)\cr
\ro u(t)^a&=e^{i t}\ro u^a,~~\rstate {a_1,\dots a_k}(t)
=e^{ik  t}\rstate {a_1,\dots a_k}
\end{eq}

The complex representation of the three position translations
$\R^3\inmap\C^3$ leads to a color $\SU(3)$-invariance - with Gell-Mann matrices 
 \begin{eq}{l}
\chi_A\la^A={\scriptsize\pmatrix{
\chi_3+{\chi_8\over\sqrt3}&\chi_1-i\chi_2&\chi_4-i\chi_5\cr
\chi_1+i\chi_2&-\chi_3+{\chi_8\over\sqrt3}&\chi_6-i\chi_7\cr
\chi_4+i\chi_5&\chi_6+i\chi_7&-2{\chi_8\over\sqrt3}\cr}},~~

\cl C={i\over 2}\ro u^a\la_a^b\ro u_b^\star,~~[\cl C,H]=0
\end{eq}The $\SU(3)$-representations $[2C_1,2C_2]$ are characterized
by two integers, they have the dimension
\begin{eq}{l}
\dim_\C[2C_1,2C_2]=(1+2C_1)(1+2C_2)(1+C_1+C_2)
\end{eq}The harmonic oscillator representations $[k,0]$
(singlet, triplet, sextet, etc.) 
are the totally symmetric  products  
 of $\SU(3)$-triplets $[1,0]$  
\begin{eq}{l}
{\OD^{k}}\C^3\cong\C^{{2+k\choose2}},~~\dim_\C[k,0]={2+k\choose k}=1,3,6,\dots,~~k=0,1,2,\dots\cr
\end{eq}

The  rotation group, generated by
the transposition antisymmetric Lie subalgebra 
\begin{eq}{l}

\SO(3)\inmap \SU(3)\hbox{ with }
\left\{\begin{array}{l}
\chi_Ai{\la^A-(\la^A)^T\over 2}
={\scriptsize\pmatrix{
0&\chi_2&\chi_5\cr
-\chi_2&0&\chi_7\cr
-\chi_5&-\chi_7&0\cr}}
={\scriptsize\pmatrix{
0&\phi_3&-\phi_2\cr
-\phi_3&0&\phi_1\cr
\phi_2&-\phi_1&0\cr}}
\cr
\cl L^a=\ep^{abc}\ro u^b\ro u^\star_c,~~
(\cl L^1,\cl L^2,\cl L^3)=(\cl C^7,-\cl C^5,\cl C^2)\end{array}\right.
\end{eq}comes with the real 5-dimensional
orientation manifold, i.e. the rotation group orbits $\SU(3)/\SO(3)$
in the color group. which describes the ${4\choose2}-1$
relative phases  of the three position
directions in complex quantum structures. 
The principal 
quantum number $k=L+2N$
is the sum of the angular momentum quantum number 
$L$  for $\SO(3)$ and the radial quantum number (knot number)
$N$ for the rotation group classes in 
$\SU(3)\cong\SO(3)\x \SU(3)/\SO(3)$.
One has with the angular momentum degeneracy
$1+2L$ the energy degeneracy
given by the dimensions of  the $\SU(3)$-representations
\begin{eq}{l}
\hbox{multiplicity }{k+2\choose2},~~
[k,0]\stackrel{\SO(3)}\cong
\left\{\begin{array}{ll}
{\PL_{L=0}^k}[2L]
,&k=0,2,\dots\hbox{ (even)}\cr
{\PL_{L=1}^k}[2L]
,&k=1,3,\dots\hbox{ (odd)}\cr\end{array}\right.\cr

E_{LN}-{3\over2}=L+2N=k 
\then (L,N)=\left\{\begin{array}{l}
(k,0),(k-2,1),\dots,(0,{k\over2})\cr
(k,0),(k-2,1),\dots,(1,{k-1\over2})
\end{array}\right.\cr
\end{eq}

\chapter{Free Particles as Translation Orbits}

Free particle properties are determined by the Poincar\'e group and its
irreducible Hilbert representations\cite{WIG}.
Compact time re\-pre\-sen\-ta\-tion
matrix elements $t\mape e^{iEt}$, multiplied either with 
compact   position re\-pre\-sen\-ta\-tion matrix elements,
e.g.
$(t,\rvec x)\mape e^{iEt}~e^{-i\rvec q\rvec x}$ and 
$(t,\rvec x)\mape e^{iEt}~{\sin|\rvec q|r\over r}$,
or with 
 noncompact ones, e.g.  $(t,\rvec x)\mape e^{iEt}~e^{-|Q|r}$,  
are Lorentz compatibly connected in 
Feynman propagators of particle fields for position dimension $s=1,2,3,\dots$
\begin{eq}{l}
{ i\over\pi}\int{d^{1+s}q\over(2\pi)^s}
{1\over q^2+io-m^2} e^{iqx}\hbox{ with }
{i\over\pi}{1\over q^2+io-m^2}
=\de(q^2-m^2)+{i\over\pi}{1\over q_\ro P^2-m^2}
\end{eq}They involve  as energy-momentum measures  the Dirac distribution  and 
the principal value pole distribution with $q_\ro P^2$.

\section{Re\-pre\-sen\-ta\-tions of  Cartan Spacetime}

2-dimensional spacetime translations  without rotation degrees of freedom
are  acted upon with the  orthochronous abelian 
Lorentz group - they constitute  Cartan spacetime
\begin{eq}{rl}
\SO_0(1,1)\sx\R^2
\end{eq}

The nonrelativistic 
free scattering waves are embedded 
with compact translation 
re\-pre\-sen\-ta\-tions $x\mape e^{iqx}$, integrated with 
$\SO_0(1,1)$-invariant measures of the  momenta 
 or of the kinetic energies over the threshold
\begin{eq}{l} 
\int d^2q~\de(q_0^2-q_3^2-m^2)\then\left\{\begin{array}{ll}
{dq_3\over q_0}
&\hbox{with }q_0=\sqrt{q_3^2+m^2}\cr
{dq_0\vth(q_0^2-m^2)\over q_3}&\hbox{with }q_3=\sqrt{q_0^2-m^2}\cr\end{array}\right.
\end{eq}This leads to the on-shell
propagator contribution with compact re\-pre\-sen\-ta\-tion matrix elements
of the translation group $\R^2$
\begin{eq}{r}
\R^2\ni x=(t,z)\mape \int d^2q~\de(q^2-m^2)e^{iqx}
=\left\{\begin{array}{l}
\int{dq_3\over q_0}~
e^{-iq_3z}~\cos q_0t\cr
\int {dq_0\vth(q_0^2-m^2)\over |q_3|}~e^{iq_0t}~\cos q_3 z\end{array}\right.\cr
\end{eq}

The indefinite unitary (noncompact)
 re\-pre\-sen\-ta\-tions (`off shell') of the translations 
  embed the bound waves. They
come in   Fourier transformed principal value distributions of energy-momenta
and are integrated with the 
invariant  measure for binding energies below the threshold
\begin{eq}{l}
{dq_0\vth(m^2-q_0^2)\over |Q|}\hbox{ with }|Q|=\sqrt{m^2-q_0^2}
\end{eq}There arise both  contributions from compact
and noncompact position  re\-pre\-sen\-ta\-tions
(`on shell' and `off shell' with real and imaginary `momenta' resp.)
\begin{eq}{rl}
\R^2\ni x&\mape 
\int d^2q ~{1\over q_\ro P^2-m^2}e^{iqx}\cr
&=\left\{\begin{array}{l}
-\pi\int {dq_3\over q_0}~e^{-iq_3z} \sin q_0|t|
\cr
 \pi\int dq_0e^{iq_0t}[{\vth(q_0^2-m^2)\over |q_3|}~\sin|q_3 z|
-{\vth(m^2-q_0^2)\over |Q|}~ e^{-|Qz|}]\cr
\end{array}\right. 
\end{eq}

\section{Re\-pre\-sen\-ta\-tions of Minkowski Spacetime}

4-dimensional  spacetime translations\cite{S96} 
with  nontrivial rotation degrees of freedom
 come with the action of the orthochronous 
Lorentz group
\begin{eq}{rl}
\SO_0(1,3)\sx\R^4
\end{eq}The Lorentz invariant measures for momenta and energies (kinetic and
binding) involve the 2-sphere measure 
$\int d\Om^2=\int_0^{2\pi}d\phi\int_{-1}^1d\cos\th$
\begin{eq}{rl} 
{d^3q\over q_0}={q^2\vth(q)dq \over q_0}~ d\Om^2
&\hbox{with }q_0=\sqrt{\rvec q^2+m^2}\cr
dq_0\vth(q_0^2-m^2) |\rvec q|~d\Om^2&
\hbox{with }|\rvec q|=\sqrt{q_0^2-m^2}\cr
dq_0\vth(m^2-q_0^2) |Q|~d\Om^2
&\hbox{with }|Q|=\sqrt{m^2-q_0^2}\cr
\end{eq}

The
$(r=0)$ regular  spherical Bessel 
function $j_0(r)={\sin r\over r}$ 
is a  compact position re\-pre\-sen\-ta\-tion matrix element
spread over the  2-sphere with the Kepler  factor ${1\over r}$
\begin{eq}{l}
\R^4\ni x=(t,\rvec x )\mape \int d^4q~\de(q^2-m^2)e^{iqx}
=\left\{\begin{array}{l}
\int {d^3q\over  q_0}~e^{-i\rvec q\rvec x}~\cos q_0t\cr
2\pi \int dq_0\vth(q_0^2-m^2)~e^{iq_0t}~
{\sin |\rvec q|r\over r}\cr
\end{array}\right.\cr
\end{eq}

For the indefinite unitary (noncompact) position re\-pre\-sen\-ta\-tion
(imaginary `momenta')
the additional rotation degrees of freedom for the nonabelian 
group $\SO(3)$ 
change  the situation  in comparison to Cartan spacetime:
The simple principal value pole does not yield
a position re\-pre\-sen\-ta\-tion matrix element, as seen,
e.g., in the $(r=0)$-singular Yukawa potential  ${e^{-|Q|r}\over r}$
\begin{eq}{l}
\R^4\ni x\mape 
-\int d^4q~{1\over q_\ro P^2-m^2}e^{iqx}
=\left\{\begin{array}{l}

\pi\int {d^3q\over  q_0}~e^{-i\rvec q\rvec x}\sin q_0|t|\cr
 2\pi^2\int dq_0 e^{iq_0t}{\vth(q_0^2-m^2)\cos|\rvec q|r
 + \vth(m^2-q_0^2)e^{-|Q|r}\over r}\cr
\end{array}\right. 

\end{eq}Position re\-pre\-sen\-ta\-tion matrix elements have to
be regular 
at $r=0$. They start with a dipole 
\begin{eq}{l}
\R^4\ni x\mape 
\int d^4q~{1\over (q_\ro P^2-m^2)^2}e^{iqx}
=\left\{\begin{array}{ll}
-\pi\int {d^3q\over  q_0}~e^{-i\rvec q\rvec x}
~{\sin q_0|t|-q_0|t|\cos q_0t\over 2 q_0^2}\cr
\pi^2\int dq_0 e^{iq_0t}[{\vth(q_0^2-m^2)\over |\rvec q|}\sin|\rvec q|r
+{\vth(m^2-q_0^2)\over |Q|} e^{-|Q|r}]\cr
\cr
\end{array}\right. 

\end{eq}

Dipoles do not occur in the Feynman propagators for
particles. The connection between spacetime dimensionality and
higher order poles
will be discussed in the overnext section.

\section{Free Particles as Orbits of Flat Spacetime}

Free particles, as defined by Wigner\cite{WIG},  are acted upon with 
Hilbert space re\-pre\-sen\-ta\-tion 
of
the semidirect (covering) Poincar\'e group 
\begin{eq}{l}
\hbox{sym\-met\-ry for free particle fields: }
\U(1)\x\SL(\C^2)\sx\R^4
\end{eq}with the homogeneous internal  phase group $\U(1)$, 
e.g. the electromagnetic group, the Lorentz (cover) group 
$\SL(\C^2)$ and the inhomogeneous 
spacetime translations.
They have a translation invariant
$m^2$ (mass),  a  rotation invariant $J$ ($\SU(2)$-spin for $m^2>0$
or $\SO(2)$-polarization for $m^2=0$) and an internal phase number, 
e.g. an electromagnetic charge number.
These properties
determine the Poincar\'e group re\-pre\-sen\-ta\-tions
\begin{eq}{l}
\int d^4q~(q)^{2J}\de(q^2-m^2)e^{iqx}
\hbox{ with }J\in\{0,{1\over 2},1,\dots\}\cr
\end{eq}For nontrivial angular momentum, 
mass and momentum have to be embedded in
Lorentz vectors, starting with the two Weyl re\-pre\-sen\-ta\-tions, left and right
handed
\begin{eq}{l}
(m,\pm \rvec q)
\inmap 
(q,\hat q)=q_0\pm\rvec q={\scriptsize\pmatrix{q_0\pm q_3&\pm(q_1-iq_2)\cr 
\pm(q_1+iq_2)&q_0\mp q_3\cr}}
\hbox{ with }q^2=m^2
\end{eq}

The re\-pre\-sen\-ta\-tion matrix elements of the Poincar\'e group  are
 Fock state expectation
values, e.g. of a free scalar Bose field $\bl\Phi$ (like a neutral stable pion)
or of a
free massive spin ${1\over2}$ Fermi field $\bl\Psi$ (like an electron-positron)
\begin{eq}{cll}
\angle{\acom{\bl\Phi(z)}{\bl\Phi(y)}}&=
 \int {d^4q\over(2\pi)^3}\hfill m~\de(q^2-m^2)e^{iqx}&\hbox{ with }x=y-z\cr
\angle{[\bl\Psi(z),\ol{\bl\Psi}(y)]}&=
 \int {d^4q\over(2\pi)^3}(\ga q+m)\de(q^2-m^2)e^{iqx}\cr
\end{eq}

The irreducible  Hilbert space  re\-pre\-sen\-ta\-tions 
of the Poincar\'e group 
are induced\cite{WIG} 
(more below for the general inducing procedure) by 
re\-pre\-sen\-ta\-tions of the direct product subgroups
with the compact fixgroups (`little groups') for the energy-momenta
\begin{eq}{l}
\hbox{sym\-met\-ry of free particles: }
\left\{\begin{array}{ll}
\U(1)\x \SU(2)\x\R^4,&m^2>0\cr
\U(1)\x\SO(2)\x \R^4,&m^2=0\end{array}\right.
\end{eq}The groups act upon the 
creation and an\-ni\-hi\-la\-tion operators
for particles $(\ro u(\rvec q),\ro u^\star(\rvec q))$, e.g. 
 for the neutral pion field as 
 the direct integral
 with  Lorentz invariant measure of the boosts 
$\SL(\C^2)/\SU(2)\cong\cl Y^3\cong \R^3$
\begin{eq}{l}
\bl \Phi=\sqrt{2m} \plintq3
 { \ro u(\rvec q)+ \ro u^\star(\rvec q)\over\sqrt2}\hbox{ with }
 q_0=\sqrt{m^2+\rvec q^2}\cr
\end{eq}For nontrivial rotation properties the
creation and an\-ni\-hi\-la\-tion operators with rotation group action
are embedded into the particle
fields with Lorentz group action
by re\-pre\-sen\-ta\-tions of the sym\-met\-ric spaces involved, i.e.
the boosts $\SL(\C^2)/\SU(2)\cong\cl Y^3$ 
for massive and  $\SL(\C^2)/\SO(2)\cong\cl Y^3\x\Om^2$ for massless particles.
E.g.  the electron with
antiparticles $(\ro a(\rvec q),\ro a^\star(\rvec q))$
(positron) is acted upon with  the left and right-handed 
Weyl re\-pre\-sen\-ta\-tions $(s,\hat s)$ of the boosts
\begin{eq}{rl}
\bl\Psi&=
{\scriptsize\pmatrix{
 \bl l^{\dot A}\cr
\bl  r^{A}\cr}}= \sqrt{2m}\plintq3
{\scriptsize\pmatrix{s({q\over m})^{\dot A}_a
 {\ro u(\rvec q)^a+\ro a^\star(\rvec q)^ a\over\sqrt2}\cr
\hat s({q\over m})_a^{ A}
{ \ro u(\rvec q)^a- \ro a^\star(\rvec q)^a\over\sqrt2}\cr}}\cr
\hbox{ with}&
\left\{\begin{array}{rl}
(s,\hat s)({q\over m})=&
\sqrt{{m+q_0\over 2m}}(\bl 1_2
\pm{\rvec q\over m+q_0})\cr
&=e^{\pm{\rvec\psi\over2}},~
\rvec\psi={\rvec q\over |\rvec q|}{\rm artanh}{|\rvec q|\over q_0}\cr

\hbox{spin }\SU(2):&a=1,2\cr
\hbox{Lorentz }\SL(\C^2):&A,\dot A=1,2\end{array}\right.\cr
\end{eq}The  translation orbits 
give the free particle fields on  spacetime, e.g.
\begin{eq}{l}
\R^4\ni x\mape \bl\Phi(x)=\sqrt{2m} \plintq3
{ e^{ipx}\ro u(\rvec q)+  e^{-ipx}\ro u^\star(\rvec q)\over\sqrt2}
\end{eq}

With the exception of 
an abelian phase  group $\U(1)$,
free particles have only spacetime related sym\-met\-ries.
Strictly speaking, free particles have
no color sym\-met\-ry, no
isopin sym\-met\-ry\cite{S003}. This is seen, e.g. for isospin, in the different masses
of proton and neutron (always as free particles) which is implemented 
in the standard model of electroweak and strong interactions\cite{WEIN}
by  the ground state
degeneracy: Hyperisospin, i.e. hypercharge-isospin,
 $\U(2)$-sym\-met\-ry 
 for the interactions
is `bleached' to the electromagnetic
$\U(1)$-sym\-met\-ry for free particles; isospin  
disappears as sym\-met\-ry, the multiplicities remain.
 Color $\SU(3)$-sym\-met\-ry 
  for the strong interactions is postulated
to be even confined for  free particles, there remain only color
trivial singlets.

\section{Multipole Energy-Momentum Measures}

In Feynman propagators there arise   
measures for spheres $\Om^s\cong \SO(1+s)/\SO(s)$,
$s=1,2\dots$  and hyperboloids 
$\cl Y^s\cong \SO_0(1,s)/\SO(s)$. 
They
 can be constructed
with the defining representations of $\SO(1+s)$
and $\SO_0(1,s)$
acting on energy-momenta (imaginary `momenta' for spheres)
\begin{eq}{rl}

\log\SO(1+s)\rvec\pl\R^{1+s}:&
{\scriptsize\(\begin{array}{c|c}
0&i\rvec\chi\cr\hline
i\rvec\chi&\log\SO(s)\cr\end{array}\)}
{\scriptsize\(\begin{array}{c}
q_0\cr\hline
i\rvec q\cr\end{array}\)}\cr
&q^2=q_0^2-(i\rvec q)^2=q_0^2+\rvec
q^2\cr

\log\SO(1,s)\rvec\pl\R^{1+s}:&
{\scriptsize\(\begin{array}{c|c}
0&\rvec\psi\cr\hline
\rvec\psi&\log\SO(s)\cr\end{array}\)}
{\scriptsize\(\begin{array}{c}
q_0\cr\hline
\rvec q\cr\end{array}\)}\cr
&q^2=q_0^2-\rvec q^2\cr

\end{eq}Positive vectors $q^2>0$ have  $\SO(s)$ as fixgroup.
The representations of the 
fixgroup classes  are parametrizable by 
unit vectors   
\begin{eq}{l}
q^2=1:~~\left\{\begin{array}{rlll}

 {\scriptsize\(\begin{array}{c}
q_0\cr\hline
i\rvec q\cr\end{array}\)}&= {\scriptsize\pmatrix{\cos\chi \cr 
{\rvec q\over|\rvec q|}i\sin\chi\cr}}
,&{\scriptsize\(\begin{array}{c|c}
q_0&iq_a\cr\hline
iq_b& \de_{ab}-{q_aq_b\over 1+q_0}\cr
\end{array}\)}&\rin\SO(1+s)/\SO(s)\cr

 {\scriptsize\(\begin{array}{c}
q_0\cr\hline
\rvec q\cr\end{array}\)}&= {\scriptsize\pmatrix{\cosh\psi \cr {\rvec q\over|\rvec q|}\sinh\psi\cr}}
,&{\scriptsize\(\begin{array}{c|c}
q_0&q_a\cr\hline
q_b& \de_{ab}+{q_aq_b\over 1+q_0}\cr
\end{array}\)}&\rin\SO_0(1,s)/\SO(s)\cr
\end{array}\right.
\end{eq}Unit vectors $q^2=1$ can be used to parametrize the  positive $\SO(1+s)$ 
and $\SO_0(1,s)$-invariant measures, unique up to a factor.
In addition, there are  other parametrizations, 
both with a finite
and infinite range - with an
trigonometric angle  $\chi$  or hyperbolic `angle' $\psi$, with 
imaginary `momenta' $ip$  and real 
momenta $p$ and with imaginary Riemann parameters
$iv$ for spheres and real Poincar\'e parameters $v$ for hyperboloids.

The parametrizations for 1-sphere and 1-hyperboloid  
are
\begin{eq}{llll}
\Om^1\ni
{\scriptsize\pmatrix{\cos\chi\cr i\sin\chi\cr}}
&={\scriptsize\pmatrix{q_0\cr iq\cr}}
&={1\over \sqrt{1+p^2}}{\scriptsize\pmatrix{1\cr ip\cr}}
&={2\over 1+v^2}{\scriptsize\pmatrix{{1-v^2\over2}\cr iv\cr}}

\cr
\cl Y^1\ni
{\scriptsize\pmatrix{\cosh\psi\cr i\sinh\psi\cr}}
&={\scriptsize\pmatrix{|q_0|\cr q\cr}}
&={1\over \sqrt{1-p^2}}{\scriptsize\pmatrix{1\cr p\cr}}
&={2\over 1-v^2}{\scriptsize\pmatrix{{1+v^2\over2}\cr v\cr}}

\end{eq}Sphere parametrizations with a square root give the half sphere only.
From the 1-forms
\begin{eq}{llll}
{\scriptsize\pmatrix{-\sin\chi d\chi\cr i\cos\chi d\chi\cr}}
&={\scriptsize\pmatrix{dq_0\cr idq\cr}}
&={1\over \sqrt{1+p^2}^3}{\scriptsize\pmatrix{-pdp\cr idp\cr}}
&={1\over (1+v^2)^2}{\scriptsize\pmatrix{-4vdv\cr i2(1-v^2)dv\cr}}

\cr
{\scriptsize\pmatrix{\sinh\psi d\psi\cr \cos\psi d\psi\cr}}
&={\scriptsize\pmatrix{d|q_0|\cr dq\cr}}
&={1\over \sqrt{1-p^2}^3}{\scriptsize\pmatrix{pdp\cr dp\cr}}
&={1\over (1-v^2)^2}{\scriptsize\pmatrix{4vdv\cr2(1+v^2)d v\cr}}

\end{eq}one obtains the measures
\begin{eq}{l}
\begin{array}{llll}
\int d\Om^1&=\int_{-\pi}^\pi d\chi
&=\int_{-1}^1{dq_0\over\sqrt{1-q_0^2}}
&=\int_{-1}^1{dq\over\sqrt{1-q^2}}\cr
&&=\int_{-\infty}^\infty {2dp\over 1+p^2}
&=\int_{-\infty}^\infty {2dv\over 1+v^2}=\oint_{p=i} {2dp\over 1+p^2}=
2\pi \cr
\int d\cl Y^1&=\int_{-\infty}^\infty d\psi
&=\int_{1}^\infty {dq_0\over
\sqrt{q_0^2-1}}
&=\int_{-\infty}^\infty {dq\over
2\sqrt{1+q^2}}\cr
&&= \int_{-1}^1 {dp\over1-p^2}&= \int_{-1}^1 {dv\over1-v^2}\end{array}
\end{eq}The $v$-parametrization 
coincides with the momentum $p$-parametrization only 
for $s=1$.

The abelian case $s=1$ is embedded in  the general case
with the $(s-1)$-sphere and  factors $q^{s-1}$ with $|\rvec q|=q$ 
\begin{eq}{rl}
\int 
{\scriptsize\pmatrix{
d\Om^s\cr
d\cl Y^s\cr}}&=

\int d^{1+s} q
{\scriptsize\pmatrix{
\hfill\de(q_0^2+\rvec q ^2-1)\cr
\vth(q_0)\de(q_0^2-\rvec q ^2-1)\cr}}\cr
\cr
&=\int d\Om^{s-1}
\int dq_0
\int_0^\infty q^{s-1}dq
{\scriptsize\pmatrix{
\hfill\de(q_0^2+ q ^2-1)\cr
\vth(q_0)\de(q_0^2-q ^2-1)\cr}}
\end{eq}The $0$-sphere consists of two points
\begin{eq}{l}
\int d\Om^{s-1}=|\Om^{s-1}|=
{2\pi^{{s\over2}}\over \Ga({s\over2})}
=2, 2\pi,4\pi,2\pi^2,{8\pi^2\over 3},\dots,~~
|\Om^0|=\card\{1,-1\}
\end{eq}The momentum $p$-parametrization
is square root free
for odd position dimension $s=1,3,\dots$,  
the  $v$-parametrization
for all $s=1,2,\dots$ - for the spherical measures
\begin{eq}{rll}
\int d\Om^s&=
\int d\Om^{s-1}{1\over2}\int_{-1}^1{dq_0\over \sqrt{1-q_0^2}^{2-s}}
&=\int d\Om^{s-1}
\int_{0}^1{q^{s-1}dq\over\sqrt{1-q^2}}\cr

&=\int d\Om^{s-1}
\int_0^\infty {2p^{s-1}dp\over \sqrt{1+p ^2}^{1+s}}
&=\int d\Om^{s-1}
\int_0^\infty  v^{s-1}dv~({2\over 1+v^2})^s\cr

\end{eq}and the hyperbolic ones
\begin{eq}{rll}

\int d\cl Y^s&=
\int d\Om^{s-1}{1\over2}\int_{1}^\infty {dq_0\over \sqrt{q_0^2-1}^{2-s}}
&=
\int d\Om^{s-1}
{1\over2}\int_{0}^\infty {q^{s-1}dq\over\sqrt{1+q^2}}\cr
&=\int d\Om^{s-1}
\int_{0}^1{p^{s-1}dp\over \sqrt{1-p ^2}^{1+s}}
&=\int d\Om^{s-1}
{1\over2}\int_{0}^{1}  v^{s-1}dv~({2\over 1-v^2})^s\cr
\end{eq}summarized as follows
\begin{eq}{rlll}
\int d\Om^s&=
\int_{\rvec q^2\le1}{d^sq\over\sqrt{1-\rvec q^2}}&=
\int {2d^sp\over \sqrt{1+\rvec p ^2}^{1+s}}
&=\int   d^s v~({2\over 1+\rvec v^2})^s\cr
\int d\cl Y^s&=\int {d^sq\over2\sqrt{1+\rvec q^2}}
&=
\int_{\rvec p^2\le1} {d^sp\over \sqrt{1-\rvec p ^2}^{1+s}}
&={1\over2}\int_{\rvec v^2\le1}  d^sv~({2\over 1-\rvec v^2})^s\cr

\end{eq}

The $\SO(4)$-invariant $\Om^3$-measure 
of the momenta $ d^3 p~({2\over Q^2+\rvec p^2})^2$ 
has been used above for the bound states in the 
nonrelativistic Kepler dynamics with $Q^2$ related to the bound state energy.

\chapter{Harmonic Analysis of Interactions}

The representations of interactions are different from those of free particles.
Spacetime interactions are implemented by
the off-shell part in Feynman propagators. They are supported
by the causal bicone.
The harmonic analysis 
of functions on  the future cone as a homogeneous spacetime model
displays the irreducible re\-pre\-sen\-ta\-tions of the
acting group.
Those representations of the extended Lorentz group $\GL(\C^2)$
have to be used for spacetime interactions.
For the harmonic analysis of nonlinear spacetime,
 free particle fields are not
enough. Additional genuine interaction fields with 
higher order pole energy-momentum measures are necessary\cite{HEI,S913}.

\section{The  Causal Support of Interactions}

Feynman propagators, e.g. for a neutral pion field
\begin{eq}{rl}
\angle{\acom{\bl\Phi}{\bl\Phi}(x)
-\ep(x_0)\com{\bl\Phi}{\bl\Phi}(x)}
&= { i\over\pi}\int{d^4q\over(2\pi)^3}
{m\over q^2+io-m^2} e^{iqx}\cr
&=m \int {dq_0\over i (2\pi)^2}
e^{iq_0x_0}{
\vth(q_0^2-m^2)~e^{i|\rvec q|r}
+\vth(m^2-q_0^2)~e^{-|Q|r}\over r}
\cr
\hbox{with}&
|\rvec q|=\sqrt{q_0^2-m^2},~~
|Q|=\sqrt{m^2-q_0^2}
\end{eq}are the sum of compact and noncompact
position re\-pre\-sen\-ta\-tions as seen in the 
on-shell off-shell decomposition of the energy-momentum 
measures into  free particle and interaction measure
\begin{eq}{l}
{i\over\pi}{1\over q^2+io-m^2}
=\de(q^2-m^2)+{i\over\pi}{1\over q^2_\ro P-m^2}
\end{eq}The off-shell part
$\ep(x_0)\com{\bl\Phi}{\bl\Phi}(x)$
 with the principal value integration has an 
explicit spacetime  dependence $\ep(x_0)\vth(x^2)$
which is not implemented by
particle fields
\begin{eq}{l}
 { i\over\pi}\int{  d^4q\over
   q^2_\ro P-m^2} e^{iqx}
=-\ep(x_0) \int  d^4q
~\ep(q_0)\de(q^2-m^2)e^{iqx}
\end{eq}Relativistic interaction structures, e.g.
Yukawa potentials ${e^{-|Q|r}\over r}$,
are  supported by the   causal 
spacetime bicone. They arise as  Fourier transformed principal values
\begin{eq}{rl}
-\int {d^4q\over \pi^3}
{1\over q^2_{\ro P}-m^2}e^{iqx}
=&
{\p\over\p {x^2\over 4}} \vth(x^2)
\cl J_0(\sqrt{m^2x^2})
=\de({x^2\over 4})+ \vth(x^2)
{\p\over\p {x^2\over 4}}\cl J_0(\sqrt{m^2x^2})\cr
\end{eq}involving the  Bessel function $\cl J_0$.

The advanced energy-momentum 
pole measures
\begin{eq}{rl}
{1\over i\pi}{1\over (q-io)^2-m^2}
&=\ep(q_0)\de(q^2-m^2)+{1\over i\pi}{1\over q^2_\ro P-m^2}\cr
&\hbox{with }(q-io)^2=(q_0-io)^2-\rvec q^2\cr

\end{eq}have as Fourier transforms  distributions with  future support
\begin{eq}{l}
\int {d^4q\over \pi^3}
{1\over (q-io)^2-m^2}e^{iqx}
=-
\vth(x_0){\p\over\p {x^2\over 4}}
 \vth(x^2)
\cl J_0(\sqrt{m^2x^2})
\end{eq}

Free particles represent the  spacetime translations $\R^4$,
interactions  the causal bicone $\R^4_\od\uplus\R^4_\and$,
future and past.

\section{Linear and Nonlinear Spacetime}

Minkowski translations $\R^4$ (4-di\-men\-sio\-nal) 
contain, on the one side,  
position translations 
$\R^3$ (3-di\-men\-sio\-nal)
 and, on the other side, 
 Cartan translations $\R^2$ 
(2-di\-men\-sio\-nal) with the 1-di\-men\-sio\-nal time and position translations $\R$ resp.
\begin{eq}{ccccc}
\tau\in\R
&\map &x^0+\si_3x^3\in\R^2&\map&
x=x^0+\rvec x
={\scriptsize\pmatrix{x^0+x^3&x^1-ix^2\cr x^1+ix^2&x^0-x^3\cr}}\in\R^4
\cr
&\nearrow&&\nearrow&\cr
z\in\R&\map&\rvec x\in\R^3&&\cr
\end{eq}Cartan and Minkowski translations are parametrized by
hermitian $(2\x 2)$-matrices, $x^0\cong x^0\bl1_2$ and  
$\rvec x=x^a\si_a$, diagonal for Cartan spacetime. 
The group $\D(1)=\exp\R$ acts upon the 1-di\-men\-sio\-nal  translations  
(time and position). For the Cartan translations, it is rearranged to
$\D(1)\x\D(1)\cong\D(1)\x\SO_0(1,1)$ with the rotation free orthochronous
Lorentz group 
and then embedded
(not unique), together with the position rotations $\SO(3)$, into the
$\D(1)$-extended Poincar\'e group
\begin{eq}{ccccc}
\D(1)\sx \R
&\map &[\D(1)\x \SO_0(1,1)]\sx\R^2&\map &[\D(1)\x \SO_0(1,3)]\sx\R^4
\cr

&\nearrow&&\nearrow&\cr
\D(1)\sx \R&\map&[\D(1)\x \SO(3)]\sx\R^3
\end{eq}

Time future is embedded
 into  Cartan and Minkowski future
\begin{eq}{cccl}
\R_\od\ni \tau_\od =\vth(\tau)\tau&\inmap &
\vth(x^0)\vth(x^2)(x^0+\si_3x^3)
&=x_\od\in \R^2_\od\cr
&\inmap&
\vth(x^0)\vth(x^2)(x^0+\rvec x)
&=x_\od\in \R^4_\od
\end{eq}The futures  are used as  noncompact spaces (open
cones without `skin'), i.e. without the strict presence $x=0$ and without  lightlike
translations for nontrivial position $s=1,3$
\begin{eq}{l}
x_\od\in\R_\od^{1+s}\then x_\od^2>0,~s=0,1,3
\end{eq}

Time future is the causal group $\D(1)=\exp \R$
\begin{eq}{rl}
\R_\od\ni \tau_\od&=e^{t}\in\D(1)\cong\GL(\C)/\U(1)\cr
\end{eq}Cartan future  is the direct product of causal group and 
abelian  Lorentz group 
 \begin{eq}{l}
x_\od
=e^{\psi_0+\si_3\psi}
\in \R^2_\od\cong\D(1)\x\SO_0(1,1)\cr
\end{eq}The action of the full  linear group, called extended Lorentz group, on 
Minkowski translations
\begin{eq}{l}
\GL(\C^2)\x \R^4\map \R^4,~~g\m x= g\o x\o g^\star
\end{eq}leaves the  future invariant. This gives the orbit pa\-ra\-met\-ri\-zation
with Lie  algebra coefficients
\begin{eq}{rl}
x=e^{\psi_0+\rvec\psi}=&
u({\rvec\psi\over \psi})\o
e^{\psi_0+\si_3\psi}\o u({\rvec\psi\over \psi})^\star
\in\R^4_\od\cong\GL(\C^2)/\U(2)\cr
\hbox{with}& \left\{\begin{array}{rl}
 e^{\psi_0}&=\sqrt{x^2}\in\D(1)\cr
  e^{\pm \psi}&
  =\sqrt{{x_0\pm r\over x_0\mp r}}\in\SO_0(1,1),~~ 
|\rvec\psi|=\psi,~~ {\rvec\psi\over \psi}={\rvec x\over r}\cr
\end{array}\right.

\end{eq}There are
 two rotation degrees of freedom for the 2-sphere
$\Om^2\cong \SU(2)/\SO(2)$
\begin{eq}{l}
u({\rvec\psi\over \psi})
={\scriptsize\pmatrix{
\cos{\th\over2}&ie^{-i\phi}\sin{\th\over2}\cr
ie^{i\phi}\sin{\th\over2}&\cos{\th\over2}\cr}}
={1\over \sqrt{2r(r+x_3)}}{\scriptsize\pmatrix{
r+x_3&-ix_1-x_2\cr ix_1-x_2&r+x_3\cr}}
\rin\SU(2)/\SO(2)
\end{eq}

The futures are  irreducible orbits of $\D(1)\x\SO_0(1,s)$.
1-di\-men\-sio\-nal 
and 4-di\-men\-sio\-nal future  
are the first two entries 
in the symmetric space chain $\GL(\C^n)/\U(n)$, $n=1,2,\dots$, 
which are the
manifolds of the unitary groups in the general linear group,
canonically pa\-ra\-met\-rized in the polar decomposition $g=u\o |g|$
with the real $n^2$-di\-men\-sio\-nal ordered absolute values $x_\od=|g|=\sqrt{g^\star\o g}
\in\R^{n^2}_\od$
of the general linear group. They are the positive cone of the
ordered C*-algebras with the complex $n\x n$-matrices.

The   tangent space of future at each point  is a full  translation space 
\begin{eq}{l}
\R^{1+s}
\cong\left\{\begin{array}{ll}
\log\D(1),&s=0\cr
\log \D(1)\pl\log\SO_0(1,1),&s=1\cr
\log \GL(\C^2)/\U(2),&s=3\cr
\end{array}\right.
\end{eq}

The Lie algebra coefficients
$(\psi_0,\rvec\psi)\in\R\x \R^3$
are related to tangent time and position as follows
\begin{eq}{l}
x_\od= x_0+\rvec x=e^{\psi_0}(\cosh\psi+{\rvec\psi\over \psi}\sinh\psi)
=1+{\psi_0}+\rvec\psi+\dots
\end{eq}

Minkowski future contains many familiar homogeneous subspaces
in the manifold decomposition
\begin{eq}{l}
\D(2)\cong\D(1)\x\SL(\C^2)/\SU(2)
\end{eq}1-di\-men\-sio\-nal future $\D(1)$
is multiplied with the 3-di\-men\-sio\-nal Lobatchevski space,
i.e. with the 3-hyperboloid
\begin{eq}{rl}
{x\over \sqrt{x^2}}=e^{\rvec\psi}\in\cl Y^3&
\cong\SL(\C^2)/\SU(2)\cong \SO_0(1,3)/\SO(3)\cr
&\cong \cl Y^2\x\Om^1\cong\cl Y^1\x\Om^2
\end{eq}which contains  2-di\-men\-sio\-nal non-Euclidean planes
(2-hyperboloids)
\begin{eq}{l}
\cl Y^2\cong \SU(1,1)/\SO(2)\cong \SO_0(1,2)/\SO(2)
\end{eq}and  1-di\-men\-sio\-nal hyperboloids (abelian Lorentz groups)
 $\cl Y^1\cong\SO_0(1,1)$.
The 2-sphere and 2-hyperboloid  
are $\SO(2)$-orbits of $\Om^1$ and $\cl Y^1$ 
for the axial rotations $o(\phi)$ 
\begin{eq}{l}
\cl Y^2\cong \cl Y^1\x\Om^1,~~
\Om^2\cong \Om^1\x\Om^1\cr
{\scriptsize\pmatrix{
\cosh{\psi\over2}&e^{-i\phi}\sinh{\psi\over2}\cr
e^{i\phi}\sinh{\psi\over2}&\cosh{\psi\over2}\cr}}
=o(\phi)\o e^{\si_3{\psi\over2}}\o o^\star(\phi)\cr
 o(\phi)={1\over\sqrt2}{\scriptsize\pmatrix{
1&e^{-i\phi}\cr
-e^{i\phi}&1\cr}}\in\SU(2),~~\left\{\begin{array}{rl}
e^{\si_3{\psi\over2}}&\in\SO_0(1,1)\cr
e^{i\si_3{\chi\over2}}&\in\SO(2) \hbox{ for }\psi=i\chi
\end{array}\right.

\end{eq}Here, and in general, the decompositions and isomorphies are with
respect to the manifold structure, not 
with respect to the symmetric space structure, e.g.,
$\Om^2$ is symmetric  with respect to $\SO(3)$, $\Om^1$ only with respect to
$\SO(2)$.

\section{Harmonic Analysis of Spacetime}

Harmonic analysis of a sym\-met\-ric space $G/H$ 
with Lie real groups $G\sup H$ 
analyzes the 
square integrable functions $f\in L^2(G/H)$
with respect to a  decomposition into
irreducible re\-pre\-sen\-ta\-tion spaces 
of $G$ with the related eigenvalues and invariants.  
Hilbert spaces with irreducible faithful representations
of noncompact groups are infinite di\-men\-sio\-nal.

Harmonic analysis of the tangent groups for linear spacetime
\begin{eq}{ccccc}
\R&\map&\SO_0(1,1)\sx\R^2 &\map&\SO_0(1,3)\sx\R^4\cr
&\nearrow&&\nearrow&\cr
\R&\map&\SO(3)\sx\R^3&&
\end{eq}involves the irreducible Hilbert re\-pre\-sen\-ta\-tions for
 free scattering states and free particles (Wigner), i.e.
 the square integrable functions on affine groups (tangent groups)
 $L^2(\SO(s)\sx\R^s)$
 and  $L^2(\SO_0(1,s)\sx\R^{1+s})$, $s\ge1$.
 
Harmonic analysis of nonlinear spacetime $\D(2)\cong\D(1)\x\cl Y^3$
\begin{eq}{ccccc}
\D(1)&\map&\D(1)\x\cl Y^1&\map&\D(2)\cr
&\nearrow&&\nearrow&\cr
\cl Y^1&\map&\cl Y^3&&
\end{eq}embeds the harmonic analysis of the following flat, spherical and hyperbolic
symmetric subspaces
\begin{eq}{l}

\begin{array}{|c|c|c|c|}\hline
\dim_\R&\hbox{flat}&\hbox{spherical}&\hbox{hyperbolic}\cr\hline\hline
1&\D(1)\cong\R&\Om^1\cong\SO(2)\hfill&\cl Y^1\cong\SO_0(1,1)\hfill\cr\hline
2&&\Om^2\cong\SO(3)/\SO(2)&\cl Y ^2\cong\SO_0(1,2)/\SO(2)\cr\hline
3&&&\cl Y^3\cong\SO_0(1,3)/\SO(3)\cr\hline
\end{array}
\end{eq}

The eigenvalues in a group $G$ re\-pre\-sen\-ta\-tion are
linear forms $ (\log G)^T$ of the Lie algebra $\log G$, for spacetime called 
angular momenta and spin in the case of spherical spaces and  
energies  and momenta for flat and hyperbolic spaces.  
The Lie algebra $\log G$ 
is acted upon with  the adjoint  re\-pre\-sen\-ta\-tion of the
group $G$, its forms $\log G^T$ with the  coadjoint (dual) one. 
Invariants are multilinear
Lie algebra  forms, e.g. the bilinear Killing form.

The tangent spaces of $G/H$, all isomorphic to each other,
are isomorphic to the corresponding Lie algebra classes
 and denoted by
$\log G/H=\log G/\log H$
with $\dim_\R\log G/H=\dim_\R G-\dim_\R H$. They inherit
the adjoint action
of the group $G$, equally  its linear forms.

Harmonic analysis of all the   homogeneous spaces above 
is well known \cite{WIG, MACK,GEL5,STRICH,HEL2,FOL}
 and will be shortly repeated.

\subsection{The Abelian Groups}

For the axial rotations $\SO(2)\cong\Om^1$ with rank 1
Lie algebra $\log \SO(2)\cong\R$   there are the Fourier series
with the winding numbers $l\in\Z$  as invariants. They characterize the
irreducible re\-pre\-sen\-ta\-tions 
of the circle as integer powers 
of the defining  re\-pre\-sen\-ta\-tion 
\begin{eq}{rcl}
e^{\pm i\chi}\cong  {\scriptsize\pmatrix{
\cos\chi\cr
i\sin\chi\cr}}\in\Om^1&\lrmap& l\in \Z\cr
\hbox{irreducible re\-pre\-sen\-ta\-tions: }
e^{\pm i\chi}&\mape& (e^{i\chi})^l\cr
f(\chi)={\SUM_{l=-\infty}^\infty}e^{i l\chi}\tilde f(l)
&\lrmap&
\tilde f(l)=\int_{-\pi}^\pi{d\chi\over2\pi}~ e^{-il\chi}f(\chi)\cr
\end{eq}

Harmonic (Fourier)  analysis of the causal group $\D(1)\cong\R$
has  linear invariants (energies) in a continuous spectrum 
$ E \in[\log\D(1)]^T\cong\R$
\begin{eq}{rcl}
e^{t }\in\D(1)&\lrmap&  E  \in\R\cr
\hbox{irreducible re\-pre\-sen\-ta\-tions: }e^t&\mape&
(e^t)^{i E }= e^{it E }
\cr
f(t )=\int {d E \over2\pi} e^{i Et }\tilde f( E )&\lrmap&
\tilde f( E )=\int dt ~e^{-i E t }f(t )\cr
\end{eq}The irreducible Hilbert  re\-pre\-sen\-ta\-tions arise
by the transition to an imaginary
power of the defining  nonunitary  re\-pre\-sen\-ta\-tion 
$e^{ t }\mape (e^{ t})^i$ and, then, by a continuous real power
$e^{i t }\mape (e^{i t})^E$ of this unitary re\-pre\-sen\-ta\-tion.

Harmonic analysis of the abelian Lorentz group 
$\SO_0(1,1)\cong\cl Y^1\x\Om^0$
(1-di\-men\-sio\-nal position with $\psi=z=x_3$)
coincides with the analysis of flat space $\D(1)$ which will be rewritten
in an adapted pa\-ra\-met\-ri\-zation
\begin{eq}{rcl}
e^{\pm \psi  }\cong {\scriptsize\pmatrix{
\cosh\psi \cr
\sinh\psi \cr}}\in\cl Y^1&\lrmap& P=\pm|P|
\in\R\cr
\hbox{irreducible re\-pre\-sen\-ta\-tions: }
e^{\pm  \psi  }&\mape& (e^{\pm \psi })^{i|P|}\cr
f(\psi )=\int_0^\infty {dP\over2\pi}{\SUM_{\ep=-1}^1}
 e^{i\ep P\psi }\tilde f(|P|,\ep)&\lrmap&
\tilde f(|P|,\ep)=\int d\psi ~e^{-i\ep |P|}f(\psi )\cr
\end{eq}

The Plancherel measure of the irreducible Hilbert  
re\-pre\-sen\-ta\-tions of  $\SO(2)$ and  $\SO_0(1,1)$ 
are the counting  and Lebesque measure 
\begin{eq}{l}
{\SUM_{l\in\Z}}
={\SUM_{L=0}^\infty}~
{\SUM_{\ep=-1}^1}
\hbox{ and }
\int{dP\over 2\pi}=\int_0^\infty {dP\over2\pi}
{\SUM_{\ep=-1}^1}\cr
\end{eq} 

\subsection{Spheres}

Compact structures are analyzed with the Peter-Weyl theorem\cite{PWEYL}.
The analysis of  the 
2-sphere $\Om^2\cong\SO(3)/\SO(2)$ uses the re\-pre\-sen\-ta\-tion 
of the rotations
with rank 1 Lie algebra
$\log\SO(3)\cong\R^3$. The eigenvalues
(magnetic quantum numbers) 
are a discrete subset of a
1-di\-men\-sio\-nal tangent form subspace 
$L_3\in\Z\subnoteq\R\subnoteq[\log\SO(3)]^T\cong\R^3$.
The characterizing Casimir invariant is 
the  bilinear Killing form 
with its values $L(L+1)$ determining the angular momenta 
\begin{eq}{rcl}
{\rvec x\over r}
={\scriptsize\pmatrix{
\cos\th\cr
i\sin\th e^{\pm i\phi}\cr
}}
\in\Om^2&\lrmap& (L,L_3)\in\N\x \Z\cr
\hbox{irreducible re\-pre\-sen\-ta\-tions: }
{\rvec x\over r}&\mape&({\rvec x\over r})^L\hbox{ (traceless)}\cr
f({\rvec x\over r})={\SUM_{L=0}^\infty}~
{\SUM_{L_3=-L}^L}({\rvec x\over r})^L_{L_3}
\tilde f(L)^{L_3}
&\lrmap&
\tilde f(L)^{L_3}
=\int{d\Om^2\over4\pi}
 \ol{({\rvec x\over r})^L_{L_3}}f({\rvec x\over r})
\cr
\hbox{Plancherel measure:}&&{\SUM_{L=0}^\infty}\cr
\end{eq}The spherical harmonics as basis of the irreducible 
$\SO(3)$-re\-pre\-sen\-ta\-tion
spaces are the natural powers of
the directions in $\R^3$,  
i.e. the possible $\SO(2)$'s
in $\SO(3)$, i.e. traceless 
monomials of the defining re\-pre\-sen\-ta\-tion 
$\sqrt{4\pi\over 1+2L}\ro Y^L({\rvec x\over r})\cong({\rvec x\over r})^L$.

This structure is characteristic:
E.g., the 3-sphere $\Om^3\cong\SO(4)/\SO(3)$ has as
defining re\-pre\-sen\-ta\-tions 
unit vectors $q\in\R^4$, $q^2=1$
whereof  the  Kepler re\-pre\-sen\-ta\-tions 
$\ro Y^{(L,L)}(q)$ in the
harmonic analysis  are natural powers.

\subsection{Tangent Groups}

The irreducible Hilbert re\-pre\-sen\-ta\-tions  of the Euclidean group
in the plane $\SO(2)\sx\R^2$ 
are - for nontrivial translation invariant -
infinite di\-men\-sio\-nal.
The matrix elements 
\begin{eq}{rl}
m^2>0:&\R^2\ni\rvec x\mape
\int d^2q~ \de(\rvec q^2- m^2)e^{-i\rvec q\rvec x}
=\pi\cl J_0(  m r)\cr
&\hbox{$\SO(2)$-nontrivial with }
(\rvec q)^{L}\sim(i{\p\over\p\rvec x})^{L},~~L=0,1,\dots\cr 
&\hskip43mm{\p\over\p\rvec x}={\rvec x\over r}{\p\over \p r}=
{\rvec x\over2}{\p\over
\p {r^2\over 4}}
\end{eq}involve  integer index Bessel  functions    
\begin{eq}{rl}
\R\ni  \xi\mape\cl E_L({ \xi^2\over4})&
={\cl J_L( \xi)\over({ \xi\over2})^L}=
{\SUM_{n=0}^\infty}{ (-{ \xi^2\over4})^n\over (L+n)!n!}
=\(-{\p\over\p{ \xi^2\over 4}}\)^L\cl E_0({ \xi^2\over4})\cr
\cl E_0({ \xi^2\over4})&=\cl J_0( \xi),~~
(1+L)\cl E_{1+L}({ \xi^2\over4})
=\cl E_L({ \xi^2\over4})+{ \xi^2\over4}\cl E_{2+L}({ \xi^2\over4})
\end{eq}

The irreducible  Hilbert re\-pre\-sen\-ta\-tions  of the Euclidean group
in 3-di\-men\-sio\-nal position $\SO(3)\sx\R^3$ 
for   nontrivial translation invariant are 
induced from momentum fixgroup  $\SO(2)$ 
representations with the matrix elements
\begin{eq}{rl}
 m^2>0:&\R^3\ni\rvec x\mape
\int d^3q~{1\over2|m|}~ \de(\rvec q^2- m^2)e^{-i\rvec q\rvec x}=
  \pi j_0(  m r)= \pi {\sin  m r\over m r}\cr
&\hbox{$\SO(3)$-nontrivial with }
(\rvec q)^{L}\sim(i{\p\over\p\rvec x})^{L},~~L=2J=0,1,\dots\cr 
&\hskip43mm{\p\over\p\rvec x}={\rvec x\over r}{\p\over \p r}=
{\rvec x\over2}{\p\over
\p {r^2\over 4}}
\end{eq}They are products 
$j_L( m r)\ro P^L(\cos\th)$ 
of harmonic polynomials in one direction 
(Legendre polynomials $\ro P^L\sim\ro Y^L_0$) 
with  matching spherical (halfinteger index)  Bessel functions 
\begin{eq}{l}
\R\ni \xi\mape {2\over\sqrt\pi}{j_{L}( \xi)\over ({\xi\over2})^L}=
{\cl J_{L+{1\over2}}(\xi)\over
({\xi\over2})^{L+{1\over2}}}=
{\SUM_{n=0}^\infty}{ (-{\xi^2\over4})^n\over \Ga({3\over2}+L+n)n!}

\end{eq}

The irreducible Hilbert re\-pre\-sen\-ta\-tions  of the 
rotation free Poincar\'e group $\SO_0(1,1)\sx\R^2$ 
with Cartan translations are - for  nontrivial
translation invariant  - characterized by the
matrix elements    
\begin{eq}{rrl} 
m^2\ne0:&\R^2\ni x\mape& \int d^2q~\de(q^2-m^2)e^{iqx}
=-\pi\cl N_0(\sqrt{{m^2x^2}})\cr
&&=-\pi\vth(x^2\cl N_0(\sqrt{{m^2x^2}})
+2vth(-x^2)\cl K_0(\sqrt{-{m^2x^2}})
\cr
&\SO_0(1,1)&\hskip-3mm\hbox{-nontrivial with }
(q)^{L}\sim(-i{\p\over\p x})^{L},~~L=0,1,\dots\cr 
&&\hskip30mm{\p\over\p x}={ x\over2}{\p\over \p{x^2\over4}}

\end{eq}The
order 0 Neumann function $\cl N_0$ for real argument (timelike)
is the  Macdonald function $\cl K_0$ for imaginary argument (spacelike)
\begin{eq}{rl}
\R\ni \xi\mape
\pi\cl N_0(\xi)
&=
{\SUM_{n=0}^\infty}{ (-{\xi^2\over4})^n\over (n!)^2}
[\log{|\xi^2|\over4}+2\ga_0-2\phi(n)]=-2\cl K_0(-i\xi)\cr
\phi(0)&=0,~~\phi(n)=1+{1\over2}+\ldots +{1\over n},~~n=1,2,\dots \cr
\ga_0&=-\Ga'(1)=\lim_{n\to\infty}\brack{\phi(n)-\log n}=0.5772\ldots\cr
\end{eq}

3-di\-men\-sio\-nal Fourier transformations are related to 
1-di\-men\-sio\-nal ones by a radial derivative which produces the Kepler 
factor ${1\over r}$ 
\begin{eq}{l}
\int d^3 q~ \mu(\rvec q^2) e^{-i\rvec q\rvec x}
=-{\p\over \p {r^2\over 4\pi} }\int dq~\mu(q^2) e^{-iq r}
\end{eq}as seen for the Euclidean group above with the  2-sphere spread
of $\SO(2)$-representation matrix elements, e.g.
$2 {\sin  \xi\over \xi}=-{\p\over \p {\xi^2\over 4} }
\cos \xi$.
Equally, the Fourier transformations 
in Minkowski spacetime
are obtainable from Cartan spacetime by an  invariant derivation
(2-sphere spread)
\begin{eq}{l}
\int d^4 q~
{\scriptsize\pmatrix{1\cr\ep(q_0)\vth(q^2)\cr}}
\mu(q^2)e^{iqx}
={\p\over \p  {x^2\over4\pi}}\int d^2q~ 
{\scriptsize\pmatrix{1\cr\ep(q_0)\vth(q^2)\cr}}
\mu(q^2)e^{iqx}\Big|_{x=(x_0,r)}\cr
\end{eq}

Therewith, the irreducible 
 Hilbert  re\-pre\-sen\-ta\-tions
of the Poincar\'e group
$\SO_0(1,3)\sx\R^4$ for Minkowski translations
contain - for causal  translation invariant $m^2\ge0$ 
with the representation inducing  
fixgroups $\SU(2)$ and $\SO(2)$ - the Wigner classified particle re\-pre\-sen\-ta\-tions
with matrix elements
 \begin{eq}{rl}
m^2>0:&\R^4\ni x\mape \int d^4q~{1\over m^2}~\de(q^2-m^2)e^{iqx}=
-{\p\over\p {m^2x^2\over 4\pi}}\pi \cl N_0(\sqrt{m^2x^2})\cr
&\hbox{$\SO_0(1,3)$-nontrivial with }
(q)^{L}\sim(-i{\p\over\p x})^{L},~~L=2J=0,1,\dots\cr 
&\hskip50mm{\p\over\p x}={ x\over2}{\p\over \p{x^2\over4}}

\end{eq}For nontrivial spin invariant $J$ 
there arise the  corresponding
Lorentz compatible energy-momentum powers $(q)^{2J}$. 

\subsection{Hyperboloids}

The harmonic analysis of the abelian group $\SO_0(1,1)\cong\cl Y^1$
(1-di\-men\-sio\-nal position)  above can be 
formulated in Cartan spacetime $\R^2$ which embeds both the hyperboloid 
$\cl Y^1$ and all its
tangent spaces $\log\SO_0(1,1)\cong\R$.
With the backward-forward  momenta $\ep |P|$, $\ep=\pm1\in\Om^0$, 
as linear  forms thereon,
the defining $\SO_0(1,1)$-re\-pre\-sen\-ta\-tion can be written
 as Lorentz product
$Py=P_0y_0-P_1y_1$ with a lightlike energy-momentum $P$, i.e. $P^2=0$, 
 normalized with the momentum $|P|$  
\begin{eq}{l}
\cl Y^1\ni y= {\scriptsize\pmatrix{y_0\cr y_1\cr}}= {\scriptsize\pmatrix{
\cosh\psi  \cr
\sinh\psi \cr}},~~P=|P| {\scriptsize\pmatrix{
1\cr
\ep\cr}}\in\R_+\x\Om^0,~{Py\over|P|}=\{e^{\ep\psi }\}\cr
\end{eq}The analogue  notation in Euclidean
$\R^2$  for the defining 
 re\-pre\-sen\-ta\-tion of $\SO(2)\cong\Om^1$ reads 
with the Euclidean product $x^2=x_1^2+x_2^2$
\begin{eq}{l}
\Om^1\ni  x= {\scriptsize\pmatrix{
\cos\chi\cr
i\sin\chi\cr}},~~ l= L {\scriptsize\pmatrix{
1\cr
 \ep\cr}}\in\N\x\Om^0,~~{l x\over L }=\{e^{i\ep\chi}\}
\end{eq}

There are other pa\-ra\-met\-ri\-zations:
The $\SO_0(1,1)$-exponentials $e^{\pm\psi}$ are 
the lightcone coordinates $y_0\pm y_1$ in the tangent pa\-ra\-met\-ri\-zation
and   ${1\pm X\over 1\mp X}$ in the Poincar\'e  pa\-ra\-met\-ri\-zation
\begin{eq}{rcl}
y=
{1\over \sqrt{1-x^2}}{\scriptsize\pmatrix{
1\cr
x\cr}}={2\over 1-X^2}{\scriptsize\pmatrix{
{1+X^2\over2}\cr
X\cr}}
\in\cl Y^1&\lrmap& P=|P|{\scriptsize\pmatrix{
1\cr \pm 1\cr}}\in \R_+\x\Om^0\cr
e^{\pm \psi  }&=&y_0\pm y_1=\sqrt{1\pm x\over 1\mp x}={1\pm X\over 1\mp X}\cr
f(y)=\int {dP\over2\pi}
 \({Py\over|P|}\)^{i|P|}\tilde f(P)&\lrmap&
\tilde f(P)=\int d\cl Y^1(y)~
\({Py\over|P|}\)^{-i|P|}f(y)\cr

\end{eq}The $\cl Y^1$-measure has the corresponding pa\-ra\-met\-ri\-zations 
\begin{eq}{l}
\int d\cl Y^1=
 \int_0^\infty  
d\psi\int d\Om^0=\int{dy_1\over\sqrt{1+y_1^2}}
=\int_{-1}^1{dx\over 1-x^2}
=\int_{-1}^1{dX\over 1-X^2}
\cr
\end{eq}

The Lorentz compatible formalism with the powers of the defining 
$\SO_0(1,1)$-re\-pre\-sen\-ta\-tions 
$({Py\over |P|})^{i|P|}=(e^{\ep\psi})^{i|P|}$
as irreducible re\-pre\-sen\-ta\-tions is generalizable to higher dimensions
$\R^{1+s}$ with $\SO_0(1,s)$-invariant products:
The   hyperboloid $\cl Y^s$ (nonlinear $s$-di\-men\-sio\-nal position)
embeds  flat hyperboloids $\cl Y^1$ (1-di\-men\-sio\-nal positions), 
 related to each other by rotations
$\SO(s)$.  With $\cl Y^s$ and its tangent spaces $\R^s$ 
embedded in $\R^{1+s}$, 
the  harmonic analysis of the $\cl Y^s$-functions
use the tangent space forms $\rvec P\in\R^s$ (momenta).
The future lightcone $\ro V^s$ in $\R^{1+s}$
\begin{eq}{l}
\ro V^s\cong \SO_0(1,s)/[\SO(s-1)\sx\R^{s-1}]\cong\R_+\x\Om^{s-1}
\end{eq}is isomorphic to the $\cl Y^s$-tangent space.
The momenta $\rvec P\in(\log\cl Y^s)^T\cong\ro V^s$
can be  written as  lightlike energy-momenta  $\ro V^s=\{P\in\R^{1+s}\mid P^2=0\}$
\begin{eq}{rcl}
\R^{1+s}\supnoteq\cl Y^s\ni
y={\scriptsize\pmatrix{y_0\cr\rvec y\cr}}=
{\scriptsize\pmatrix{\cosh\psi\cr{\rvec y\over r}\sinh\psi\cr}}
&\lrmap&
\R^{1+s}\supnoteq\ro V^s\ni P=|\rvec P|{\scriptsize\pmatrix{1\cr
{\rvec P\over |\rvec P|}\cr}}
\cr
y^2=1,&&P^2=0
\end{eq}The normalized rotation $\SO(s)$-invariant product contains
the defining re\-pre\-sen\-ta\-tion of 
the hyperboloids $\SO_0(1,1)$ in $\cl Y^s$,
indexed with an $\Om^1$-orientation angle $\al$
\begin{eq}{rl}
{Py\over|\rvec P|}
&=\cosh\psi+\sinh\psi \cos \al
=e^\psi\cos^2{\al\over2}+e^{-\psi}\sin^2{\al\over2}\cr
&\hskip42.5mm
=|\cosh{\psi\over 2}+e^{-i\al}\sinh{\psi\over 2}|^2,~
\cos\al=
{\rvec P\rvec y\over |\rvec P| r}

\end{eq}The irreducible re\-pre\-sen\-ta\-tions
for the noncompact abelian subgroups $\cl Y^1(\al)\cong\SO_0(1,1)$
have  an  imaginary power
\begin{eq}{l}
\hbox{for all }\cl Y^1(\al):~~{ Py\over  |\rvec P|}\mape
(e^{\psi}\cos^2{\al\over2}+e^{-\psi}\sin^2{\al\over2})^{i|\rvec P|}
=
({ Py\over  |\rvec P|})^{i|\rvec P|}
\cr
\end{eq}Therewith, the position functions  $L^2(\cl Y^s)$ have the 
harmonic analysis
\begin{eq}{l}
f(\rvec \psi)=\int d\d{\cl Y}^s(\rvec P) 
{(e^{\psi}\cos^2{\al\over2}+e^{-\psi}\sin^2{\al\over2})|^{i|\rvec P|}\over
(e^{\psi}\cos^2{\al\over2}+e^{-\psi}\sin^2{\al\over2})^{s-1\over2}}
\tilde f(\rvec P)\cr
\hskip20mm
\lrmap
\tilde f(\rvec P)=\int d\cl Y^s(\rvec\psi) 
{
(e^{\psi}\cos^2{\al\over2}+e^{-\psi}\sin^2{\al\over2})^{-i|\rvec P|}\over
(e^{\psi}\cos^2{\al\over2}+e^{-\psi}\sin^2{\al\over2})^{s-1\over2}}
f(\rvec \psi)
\cr\end{eq}The irreducible re\-pre\-sen\-ta\-tions
are normalized 
with a length factor for the corresponding 
compact sphere $\cl Y^s\cong\cl Y^1(\al)\x \Om^{s-1}(\al)$
\begin{eq}{l}
|\cosh{\psi\over 2}+e^{-i\al}\sinh{\psi\over 2}
|^{s-1}
=({ Py\over  |\rvec P|})^{{s-1\over2}}\cr
\end{eq}

The Plancherel measure\cite{STRICH} $\Pi^s$ for 
the quadratic invariants $\rvec P^2$ of 
the arising irreducible $\SO_0(1,s)$-re\-pre\-sen\-ta\-tions reads
\begin{eq}{rl}
 \int d\d{\cl Y}^s(\rvec P) 
&={1\over(2\pi)^s}\int_0^\infty \Pi^s(\rvec P^2)
dP \int d\Om^{s-1}\cr  
\Pi^s(m^2)&=
\left|{\Ga( im+{s-1\over2})\over\Ga(i m)}\right|^2
=\left\{\begin{array}{ll}
m\tanh\pi m,&s=2\cr
m^2,&s=3\cr
\end{array}\right.
\end{eq}

E.g., the harmonic analysis  of the $\cl Y^2$-functions 
(2-dimensional hyperbolic position, non-Euclidean plane) employs the
 principal series re\-pre\-sen\-ta\-tions\cite{BARG} of the group
$\SL(\R^2)\sim\SO_0(1,2)$. The  Plancherel with $\Om^1$-measure is  
Lebesque  $d^2P$ with a hyperbolic factor
\begin{eq}{rcl}
y\in\cl Y^2\cong\cl Y^1\x\Om^1&\lrmap&
  P= |\rvec P|{\scriptsize\pmatrix{1\cr\cos\Phi\cr\sin\Phi\cr}}
 \in \ro V^2\cong \R_+\x\Om^1\cr
y={\scriptsize\pmatrix{y_0\cr y_1\cr y_2\cr}}
={\scriptsize\pmatrix{\cosh\psi \cr\sinh\psi \cos\phi\cr\sinh\psi \sin\phi\cr}}
&=&
{1\over \sqrt{1-x^2}}{\scriptsize\pmatrix{1\cr \rvec x\cr}}
={2\over 1-X^2}{\scriptsize\pmatrix{{1+X^2\over2}\cr \rvec X\cr}}\cr
{ Py\over |\rvec P|}=
\cosh\psi &\hskip-6mm+&\hskip-6mm \sinh\psi \cos(\phi-\Phi)
 =
{1+X^2+2X\cos(\phi-\Phi)\over 1-X^2}
\cr
\int d\d{\cl Y}^2(\rvec P)
&=&\int { d^2 P\over (2\pi)^2}~\tanh \pi |\rvec P|
\end{eq}Also the harmonic analysis of 
the functions on the  hyperboloid (nonlinear 3-di\-men\-sio\-nal position)
$\cl Y^3\cong\SL(\C^2)/\SU(2)$
employs  
the  principal series re\-pre\-sen\-ta\-tions\cite{NAIM} of
the Lorentz group $\SL(\C^2)\sim\SO_0(1,3)$.
Only for $s=1$ and for $s=3$ the   Plancherel 
with $\Om^{s-1}$-measure is   Lebesque $d^sP$
\begin{eq}{rcl}
y={\scriptsize\pmatrix{y_0\cr \rvec y\cr}}
\in\cl Y^3\cong\cl Y^1\x\Om^2&\lrmap&
  P= |\rvec P|{\scriptsize\pmatrix{1\cr {\rvec P\over |\rvec P|}\cr}}
 \in\ro V^3\cong\R_+\x\Om^2\cr
\int d\d{\cl Y}^3(\rvec P)
&=&\int { d^3 P\over (2\pi)^3}
\end{eq}

There is the $L^2(\cl Y^s)$-analogue 
treatment\cite{SHER} for the harmonic analysis of the
square integrable functions $L^2(\Om^s)$ on the sphere
in Euclidean $\R^{1+s}$. With
the hyperboloid-sphere transition $\psi\lrmap i\chi$
\begin{eq}{c}
\R^{1+s}\supnoteq\Om^s\ni
x={\scriptsize\pmatrix{x_0\cr i\rvec x\cr}}=
{\scriptsize\pmatrix{\cos\chi\cr i{\rvec x\over r}\sin\chi\cr}}
\lrmap
\R^{1+s}\supnoteq  l=L{\scriptsize\pmatrix{1\cr
{\rvec l\over L}\cr}},~~L=|\rvec l|
\cr
\end{eq}one  uses  the
representations of the abelian subgroups  $\Om^1(\al)\cong\SO(2)$
\begin{eq}{rl}
\hbox{for all }\Om^1(\al):~~ { lx\over L }&= 
\cos\chi+i\sin\al\sin\chi=
e^{i\chi}\cos^2{\al\over2}+e^{-i\chi}\sin^2{\al\over2}\cr
 &\mape ({lx\over L })^L 
 ,~~\cos\al={\rvec  l\rvec x\over L r}
 \end{eq}

\section
[Residual Representations of Nonlinear Spacetime]
{Residual Representations\\of Nonlinear Spacetime}

The  defining re\-pre\-sen\-ta\-tions 
of the abelian groups above
can be written as residues\cite{BESO} from Fourier transformed energy-momentum
measures\cite{S002} where the poles are at the value of the invariant 
- real for compact and  imaginary for noncompact transformations
\begin{eq}{rrl}
\U(1)\cong\Om^1:&
e^{i\chi}
=&{1\over 2i\pi}\oint {d q\over  q- 1}e^{i q\chi}=\int d q~\de( q-1)e^{i q\chi}\cr
\D(1)\cong\cl Y^1:&e^{-|\psi| }=
&{1\over \pi}\int {dq\over q^2+1}e^{-iq\psi }
\hbox{ with }\int d\Om^1=\int {2dq\over q^2+1}=2\pi
\cr
 \end{eq}This defines { basic re\-pre\-sen\-ta\-tion measures}
  - the Dirac measure for
the sphere $\Om^1$ and 
 the spherical measure for the hyperboloid $\cl Y^1$
  \begin{eq}{rrl}
\Om^1:&e^{i\chi} &\lrmap d q~\de( q-1)\cr
\cl Y^1:&e^{-|\psi |}
&\lrmap{1\over \pi}{dq\over q^2+1}={d\Om^1\over |\Om^1|}\cr
\end{eq}The measures are normalizable with a residue $1$ for the 
invariant leading to the unit for the neutral group element.

A rescaling changes the  re\-pre\-sen\-ta\-tion invariant  
\begin{eq}{l} 
 e^{il\chi}
={1\over 2i\pi}\oint {d q\over  q-l}e^{i  q\chi},~~
e^{-|Q\psi |}={1\over \pi}\int{dq|Q|\over q^2+Q^2}e^{-iq\psi }\cr
\end{eq}

With additional $\Om^2$-degrees of freedom
the rotation group is represented with  a derived Dirac measure 
\begin{eq}{l}
\SU(2)\cong \Om^1\x\Om^2:~~\left\{\begin{array}{rl}
e^{i\rvec\chi}\mape&
{1\over \pi}\int d^3 q~\de'(1-\rvec  q^2) e^{-i\rvec  q\rvec\chi}
{\scriptsize\pmatrix{
1\cr
\rvec  q\cr}}= {\scriptsize\pmatrix{
\cos|\rvec\chi|\cr -i{\rvec\chi\over |\rvec\chi|}~\sin|\rvec\chi|\cr}} \cr
\lrmap&{d^3q\over \pi}\de'(1-\rvec  q^2)\cr\end{array}\right.

\end{eq}and 
the defining Hilbert  re\-pre\-sen\-ta\-tions of Lobachevski space
with the scalar dipole measure
of the 3-sphere
and the  vector tripole measure  
\begin{eq}{l}
\cl Y^3\cong\cl Y^1\x\Om^2:~~\left\{\begin{array}{rl}
e^{\rvec\psi}\mape&
{1\over \pi^2}\int {d^3q\over (\rvec q^2+1)^2}e^{-i\rvec q\rvec\psi}
{\scriptsize\pmatrix{
1\cr
{4i\rvec q\over \rvec q^2+1}\cr}}=
{\scriptsize\pmatrix{
1\cr\rvec\psi\cr}} e^{-|\rvec\psi|}\cr
\lrmap
&{1\over \pi^2}{d^3q\over (\rvec q^2+1)^2}={d\Om^3\over |\Om^3|}
\end{array}\right.
\end{eq}

\subsection
{Representations of Cartan Spacetime}

The embedding of the defining re\-pre\-sen\-ta\-tion 
of the causal group, i.e. of 1-di\-men\-sio\-nal future,  into 
Cartan future
is given by  the Fourier transformed
advanced pole measure
\begin{eq}{rrl}
\D(1)\cong\R_\od:& \vth(\tau)&=\int {dq\over 2i\pi}~
{1\over q-io}e^{iq\tau}\cr
\D(1)\x\SO_0(1,1)\cong \R^2_\od:&\vth(x^2)\vth(x_0)&=
-\int {d^2q\over  2\pi^2}
~{1\over (q-io)^2}e^{iqx}\cr
\end{eq}Re\-pre\-sen\-ta\-tions of the position hyperboloids $\cl Y^1$ are 
embedded with an energy dependent  invariant for the $\Om^1$-measure
\begin{eq}{l}
{d^2q\over -q_\ro P^2+m_3^2}=
dq_0\vth(Q^2)~{d \Om^1(Q^2)\over2}+\dots,~~d \Om^1(Q^2)=
{2dq_3\over q_3^2+Q^2}
,~~Q^2={m_3^2-q_0^2}
\end{eq}The Hilbert re\-pre\-sen\-ta\-tions of the causal group
$\D (1)\ni e^{ t}\mape e^{i m  t}$
can be reformulated in the tangent parametrization
\begin{eq}{l}
\R_\od\ni \vth(\tau)\tau
\mape \int {dq\over2i\pi} ~{q\over (q-io)^2-m^2}e^{iq\tau}=
\vth(\tau)\cos m\tau\cr
\end{eq}

2-di\-men\-sio\-nal future 
$\D(1)\x\D(1)\cong \D(1)\x\SO_0(1,1)$ has { rank 2}. The residual re\-pre\-sen\-ta\-tions of 
these two noncompact groups are characterized by 
two invariants $(M_0^2,M_3^2)$ 
for energy and momenta, both from a continuous spectrum
\begin{eq}{rl}
\hbox{for }\D(1)\x\D(1):~~
(t,z)\mape&
\int{dq_0\over2 i\pi}{q_0\over (q_0-io)^2-M_0^2}e^{iq_0t}
\int{dq_3\over\pi}{|M_3|\over q_3^2+M_3^2}e^{-iq_3z}\cr
&=\vth(t)\cos M_0t~e^{-|M_3z|}
\end{eq}In a Lorentz compatible framework
the residual re\-pre\-sen\-ta\-tions will be  supported by 
{ two  Lorentz invariants} for 
the hyperbolic-spherical  singularity surface
with the pole function 
\begin{eq}{rl}
{1\over (q^2-m_0^2)(-q^2+m_3^2)}&={1\over m_0^2-m_3^2}
[-{1\over q^2-m_0^2}+{1\over q^2-m_3^2}]\cr
=-\int_{m_3^2}^{m_0^2}{dm^2\over m_0^2-m_3^2}~{\p\over \p m^2}{1\over q^2-m^2}
&=-\int_{m_3^2}^{m_0^2}{dm^2\over m_0^2-m_3^2}~{1\over (q^2-m^2)^2}
\end{eq}The two invariants are incorporated in a spectral
function for a dipole measure.
By the Lorentz compatible embedding
both invariants contribute to  re\-pre\-sen\-ta\-tions 
of the causal group $\D(1)$ and the position
hyperboloid $\SO_0(1,1)$.

On the lightcone $x^2=0$, where
time and position translations coincide $x^3=\pm x^0$,
the contributions  from both invariants  cancel each other
\begin{eq}{rl}
\hbox{Cartan future: }\R_\od^2\ni x_\od\mape& 
 
-{1\over\pi}\int {d^2q\over 2i\pi}{(m_0^2-m_3^2)~q~e^{iqx}\over [(q-io)^2-m_0^2][(q-io)^2-m_3^2]}\cr
=-
{\p\over\p x}[\cl E_0({m_0^2x_\od^2\over4})
-\cl E_0({m_3^2x_\od^2\over4})]
&={x_\od\over2}[m_0^2\cl E_1({m_0^2x_\od^2\over4})
-m_3^2\cl E_1({m_3^2x_\od^2\over4}]
\end{eq}

Re\-pre\-sen\-ta\-tions of future 
have -  by the corresponding tangent space integrations -
 projections\cite{S031}
 $x_\od=\tau_\od\bl1_2+z\si_3$ on re\-pre\-sen\-ta\-tions of the  causal group $\D(1)$
and of the position hyperboloids $\cl Y^1$
\begin{eq}{rrl}
\hbox{time future:}&\D(1)\cong \R_\od\ni \tau_\od\mape& 
\int {dx_3\over2\pi}
\int {d^2q\over 2i\pi}{(m_0^2-m_3^2)~q~e^{iqx}\over [(q-io)^2-m_0^2][(q-io)^2-m_3^2]}\cr

& &=\vth(\tau)[\cos m_0\tau-\cos m_3\tau]\cr

\hbox{position:}&\cl Y^1\cong \R\ni z\mape&
 \int {dx_0\over2\pi}
 \int {d^2q\over 2i\pi}{
 (m_0^2-m_3^2)~q~e^{iqx}\over [(q-io)^2-m_0^2][(q-io)^2-m_3^2]}\cr
&&={\p\over\p z}V(z)\cr
 \end{eq}The position projections 
display   exponential  interactions  
\begin{eq}{l}
V(z)={e^{-|m_0z|}\over|m_0|}-{e^{-|m_3z|}\over|m_3|},~~
{\p\over \p z}V(z)=\ep(z)(e^{-|m_0z|}-e^{-|m_3z|})
\end{eq}

\subsection
{Representations of Minkowski Spacetime}

4-di\-men\-sio\-nal future 
\begin{eq}{rrl}

\D(2)\cong \R^4_\od:& \vth(x^2)\vth(x_0)&=
\int {d^4q\over 2\pi^3}{1\over [(q-io)^2]^2}e^{iqx}
\end{eq}is the  real homogeneous space
$\D(1)\x\cl Y^3$ with rank 2 for a Cartan subgroup $\D(1)\x\SO_0(1,1)$.
Re\-pre\-sen\-ta\-tions of the position hyperboloids $\cl Y^3$ are 
embedded with an energy dependent  invariant  
\begin{eq}{rl}
{d^4q\over (q_\ro P^2-m^2)^2}
=dq_0\vth(Q^2){d\Om^3(Q^2)\over2}+\dots,~
d\Om^3(Q^2)={2d^3q \over (\rvec q^2+Q^2)^2}
\hbox{ with }Q^2={m^2-q_0^2}\cr
\end{eq}

The residual re\-pre\-sen\-ta\-tions will be  supported 
by two invariants as 
for the 2-di\-men\-sio\-nal case
with a characteristic additional dipole structure\cite{HEI} 
for the 2-sphere degrees
of freedom in 3-di\-men\-sio\-nal position   
\begin{eq}{rl}
\hbox{for }\D(1)\x[\D(1)\x\Om^2]:~~(t,\rvec x)\mape& 
\int{dq_0\over2 i\pi}{q_0\over (q_0-io)^2-M_0^2}e^{iq_0t}
\int{d^3q\over\pi^2}{|M_3|\over (\rvec q^2+M_3^2)^2}e^{-i\rvec q\rvec x}\cr
&
=\vth(t)\cos M_0t~e^{-|M_3|r}
\end{eq}In a Lorentz compatible formulation
the two invariants can be incorporated in a spectral
function for a tripole measure
\begin{eq}{rl}
{1\over (q^2-m_0^2)(-q^2+m_3^2)^2}
&={1\over (m_0^2-m_3^2)^2}[
{1\over q^2-m_0^2}-{1\over q^2-m_3^2}
-{m_0^2-m_3^2\over (q^2-m_3^2)^2}]
\cr
=\int_{m_3^2}^{m_0^2}{m_0^2-m^2\over (m_0^2-m_3^2)^2}
dm^2\({\p\over\p m^2}\)^2
~{1\over q^2-m^2}
&=2\int_{m_3^2}^{m_0^2}{m_0^2-m^2\over (m_0^2-m_3^2)^2}dm^2
~{1\over (q^2-m^2)^3}
\end{eq}The spherical invariant $m_0^2$ 
in the pole measure for $\D(1)$ and the hyperbolic invariant $m_3^2$
in the dipole measure for $\cl Y^3$ 
contribute also  to  re\-pre\-sen\-ta\-tions 
of $\cl Y^3$ and $\D(1)$.
There is one additional  continuous invariant
compared with the 
 Poincar\'e group re\-pre\-sen\-ta\-tions
for free particles.

With two invariants the vector re\-pre\-sen\-ta\-tions of 4-di\-men\-sio\-nal  future
are
\begin{eq}{l}
\hbox{Minkowski future: }
\R_\od^4\ni x_\od\mape 
{1\over\pi}\int{d^4q\over 2i \pi^2}{(m_0^2-m_3^2)^2~q~e^{iqx}\over [(q-io)^2-m_0^2][(q-io)^2-m_3^2]^2}
\cr
\hskip20mm
={\p\over\p x}\left[
{\p\over\p{x^2\over 4}}
[\cl E_0({m_0^2x_\od^2\over4})
-\cl E_0({m_3^2x_\od^2\over4})]
-(m_0^2-m_3^2)
\cl E_0({m_3^2x_\od^2\over4})\right]\cr
\hskip20mm
={x_\od\over2}[
m_0^4\cl E_2({m_0^2x_\od^2\over4})
-m_3^4\cl E_2({m_3^2x_\od^2\over4})
+(m_0^2-m_3^2)m_3^2\cl E_1({m_3^2x_\od^2\over4})]\cr

\cr
\end{eq}The projections $x_\od=\tau_\od\bl1_2+\rvec x$ on time future
and 3-di\-men\-sio\-nal position 
hyperboloids 
\begin{eq}{rrl}
\hbox{time future:}&\D(1)\cong\R_\od\ni \tau_\od\mape&
\int {d^3x\over8\pi^2}
\int{d^4q\over 2i \pi^2}{
(m_0^2-m_3^2)^2~q~e^{iqx}\over [(q-io)^2-m_0^2][(q-io)^2-m_3^2]^2}
\cr
&&\hskip-30mm=\vth(\tau)[
\cos m_0\tau-\cos m_3\tau
+{m_0^2-m_3^2\over 2m_3^2} m_3\tau\sin m_3\tau] 
\cr

\hbox{position:}&\cl Y^3\cong\R^3\ni \rvec x\mape&
\int {dx_0\over2\pi}
\int{d^4q\over 2i \pi^2}{
(m_0^2-m_3^2)^2 ~q~e^{iqx}\over [(q-io)^2-m_0^2][(q-io)^2-m_3^2]^2}

\cr&&=(m_0^2-m_3^2){\p\over\p\rvec x}V_3(r)\cr

 \end{eq}show exponential and Yukawa interactions
 - the 2-sphere spread
of a noncompact re\-pre\-sen\-ta\-tion of 1-di\-men\-sio\-nal position
with a  $z$-proportional contribution from the dipole
\begin{eq}{rl}
V_3(r)&= {e^{-  |m_3|r}\over 2|m_3|}
+{e^{-  |m_0|r}-e^{-  |m_3|r}\over(m_0^2-m_3^2) r}=
-{d\over dr^2}V_1(r)\cr
 V_1(r) &=
(1+|m_3|r){e^{-  |m_3|r}\over 2|m_3|^3}
+{{e^{-  |m_0|r}\over |m_0|}-{e^{-  |m_3|r}\over |m_3|}\over m_0^2-m_3^2}

\end{eq}

\section{Measure Convolution  and Bound States}

Feynman integrals involve convolutions of energy-momentum measures
(distributions)
\begin{eq}{l}
(\mu_1*\mu_2)(q)=\int d^4q_1d^4q_2~\mu_1(q_1)\de(q_1+q_2-q)\mu_2(q_2)
\end{eq}For  free particle energy-momentum measures
\begin{eq}{rl}
\de_{\pm|m|}(q)=\vth(\pm q_0)\de(q^2-m^2),~~m\ge0
\end{eq}the product measures display the familiar 
energy thresholds, e.g.
\begin{eq}{rl}
(\de_{\pm|m_1|}*\de_{\pm|m_2|})(q)&=2\pi\vth(\pm q_0)\vth(q^2-m_+^2)
{\sqrt{\De(q^2,m_1^2,m_2^2)}\over q^2}\cr
\De(q^2,m_1^2,m_2^2)&=(q^2-m_+^2)(q^2-m_-^2),~~
m_\pm=m_1\pm m_2
\end{eq}

In the  convolution of two Feynman particle propagators
\begin{eq}{rrl}
\pm{1\over i\pi} {1\over q^2\mp io -m_1^2}
*\pm{1\over i\pi}{1\over q^2\mp io-m_2^2}
\hskip-3mm&=2\Bigl[~~&\hskip-2mm\vth(+q_0)\de(q^2-m_1^2)
*\vth(+q_0)\de(q^2-m_2^2)\cr
&+&\hskip-2mm\vth(-q_0)\de(q^2-m_1^2)
*\vth(-q_0)\de(q^2-m_2^2)\Bigr]\cr
&\hskip5mm\pm\hfill&\hskip-5mm{1\over i\pi}
\Bigl[\de(q^2-m_1^2)*{1\over q^2_\ro P-m_2^2}+
{1\over q^2_\ro P-m_1^2}*\de(q^2-m_2^2)\Bigr]
\end{eq}the on shell-off shell convolution of the  Dirac particle measures with principal value 
interaction measures  
is not defined (`divergencies'). That is different for the 
higher order interaction measures above.

Product re\-pre\-sen\-ta\-tions can be related to measure
products as seen in the simplest example
\begin{eq}{rl}
e^{iE_1t}
e^{iE_2t}&=e^{iE_+t},~~(\de_{E_1}*\de_{E_2})(q_0)=\de_{E_+}(q_0)\cr
e^{iEt}&=\int dq_0\de(q_0-E)e^{iq_0t},~~
E_+=E_1+E_2\cr
\end{eq}Therewith bound state structures
for particles  are proposed to be
derived from pole invariants in convolution products involving the
interaction measures as described above. First steps on this
way have been tried elsewhere\cite{S031}.

\chapter{Intertwining Spacetime and Hyperisospin}

If spacetime is modeled by the equivalence classes
$\GL(\C^2)/\U(2)$, the class characterizing fixgroup $\U(2)$ arises with local
internal operations for the spacetime fields. 

The  square integrable $\C$-valued functions
$L^2(G/U)$ on a sym\-met\-ric space 
 can be rearranged  with respect to mappings of 
the fixgroup classes $G/U$ into Hilbert spaces $W$ with nontrivial re\-pre\-sen\-ta\-tions 
of the fixgroup  $U\sub G$.
A familiar example is given with the defining
re\-pre\-sen\-ta\-tion of the rotation group
\begin{eq}{l}
{\scriptsize\left(\begin{array}{c|c|c}
e^{i(\chi+\phi)}\cos^2{\th\over2}&
ie^{i\phi}{\sin \th\over\sqrt2}&
-e^{-i(\chi-\phi)}\sin^2{\th\over2}\cr
ie^{i\chi}{\sin\th\over\sqrt2}&
\cos \th&
ie^{-i\chi}{\sin\th\over\sqrt2}\cr
-e^{i(\chi-\phi)}\sin^2{\th\over2}&
ie^{-i\phi}{\sin \th\over\sqrt2}&
e^{-i(\chi+\phi)}\cos^2{\th\over2}\cr\end{array}\right)}\in\SO(3)
\end{eq}The three columns as vectors of 3-dimensional spaces are acted on with 
the defining $\SO(3)$-re\-pre\-sen\-ta\-tion.
The middle column is the spherical harmonic 
$\sqrt{4\pi\over3}\ro Y^1({\rvec x\over r})\in W_0\cong\C^3$
which, by its products,  builds up $L^2(\SO(3)/\SO(2))$ which carries a trivial 
re\-pre\-sen\-ta\-tion of the fixgroup $e^{i\chi}\in \SO(2)$.
This has been considered in the former chapter.
The left and right column with $\chi=0$ 
describe
mappings $w:\Om^2\map W_\pm $ into vector spaces $W_\pm \cong\C^3$ 
acted upon with
the nontrivial $\SO(2)$-re\-pre\-sen\-ta\-tions $\{e^{\pm i\chi}\}$.
Other examples with particle fields are given below.

The general mathematical procedure behind all this  is
the inducing of $G$-re\-pre\-sen\-ta\-tions from subgroup $U$-re\-pre\-sen\-ta\-tions
\cite{WIG,MACK,FOL}.
In nonlinear spacetime as the orientation manifold  $\D(2)\cong\GL(\C^2)/\U(2)$,
the local compact fixgroup
$\U(2)$ will be interpreted as hyperisospin.
A harmonic analysis of spacetime interactions employs
only such re\-pre\-sen\-ta\-tions of the
extended Lorentz group  
$\GL(\C^2)$ which are induced from 
re\-pre\-sen\-ta\-tions of the hyperisospin group $\U(2)$.

\section
[Dichotomy of Spacetime and Chargelike Operations]
{Dichotomy of Spacetime \\and Chargelike Operations}
 
The  real 8-dimensional linear group 
in two complex dimensions
is, via the polar decomposition
\begin{eq}{l}
\GL(\C^2)=\U(2)\o \D(2)\ni g= u(g)\o |g|,~~|g|=x_\od=\vth(x^2)\vth(x_0)x\in\D(2)
\end{eq}the  product of its  maximal compact group $\U(2)$
(hyperisospin),  
comprising the four internal or chargelike degrees of freedom,  with 
all  $\U(2)$-orbits, the  noncompact causal sym\-met\-ric space
 $\D(2)$ (nonlinear spacetime), comprising the four  external or spacetime related degrees of
freedom. 
Elementary interactions and elementary particles
are characterized by different sym\-met\-ries: 
Elementary interactions display 
 a dichotomic nonabelian sym\-met\-ry with internal (chargelike)
 and external (spacetime-like) operations 
\begin{eq}{l}
\hbox{interaction sym\-met\-ry: }\U(2)\x\GL(\C^2)\cr
\end{eq}exemplified by the subgroup product $\SU(2)\x\SU(2)$ with
isospin and spin.
The action group is a direct product of the 
hyperisospin group  and the extended
Lorentz group 
 \begin{eq}{lrll}
\hbox{internal}& \U(2)&=\U(\bl1_2)\o\SU(2)&\cong{\U(1)\x\SU(2)\over\I(2)}\cr
\hbox{external}& \GL(\C^2)
&=\D(1)\x \U(\bl1_2)\o\SL(\C^2)&\cong\D(1)\x{\U(1)\x\SL(\C^2)\over\I(2)}\cr
 \end{eq}$\I(2)=\{\pm1\}$ denotes the discrete center of the special groups,
 i.e. abelian and nonabelian transformations are centrally correlated
 \begin{eq}{l}
\U(\bl1_2)\cap\SU(2)\cong\I(2)\cong\U(\bl 1_2)\cap\SL(\C^2)
\end{eq}

In such an interpretation, the observed interaction sym\-met\-ries, e.g.
isospin and  spin $\SU(2)\x\SU(2)$, originate from one group
represented in two  different actions 
as internal and external group. 
The dichotomic interaction sym\-met\-ry
disappears in the sym\-met\-ry for free particles. In the standard model
formulation,
 the isospin $\SU(2)$-sym\-met\-ry
is `broken' (bleached), the spin $\SU(2)$-sym\-met\-ry survives\cite{S003}.

\section{Induced Re\-pre\-sen\-ta\-tions}

The doubling of a group $G$ 
to a dichotomic operation group $G\x G$  arises as a
basic  general mathematical structure: 
The binary property of the product in a group $G\x G\map G$
allows the realization
of the group on itself by both left and right multiplications, i.e. by the
realization of the square group 
\begin{eq}{l}
 G\x G\ni (u,k)\mape L_u\x R_k \hbox{ with }
L_u\o R_k (g)=ugk^{-1}\hbox{ for }g\in G
\end{eq}and subgroups  $U\x G\sub G\x G$. E.g., the diagonal subgroup
$(k,k)\in (G\x G)_\De\cong G$ is realized by the inner group automorphisms
$\Int k=L_k\o R_k$.
Or, a matrix can be acted upon from left and right - with possibly different
groups $U\x G$.
This binary structure and the related two-sided regular re\-pre\-sen\-ta\-tion,
of special interest for nonabelian operations, $\Int k(g)=kgk^{-1}\ne g$,  
is  used in the method of $G$-re\-pre\-sen\-ta\-tions induced from subgroup
$U$-re\-pre\-sen\-ta\-tions.

The inducing procedure uses a factorization\footnote{\scriptsize
Obviously, a factorization with right or  with left  cosets
is equivalent,  $U\x U\lq G\cong G\cong  G/U\x U$.}
 $G=U\x U\lq G$ with subgroup
 $U$-classes.
In the spacetime application the extended Lorentz group $G=\GL(\C^2)$
is factorized into $\U(2)\x\D(2)$ 
with the hyperisospin subgroup $U=\U(2)$ and the future cone $\D(2)$.
On the factorization $U\x U\lq G$ one constructs
re\-pre\-sen\-ta\-tions
of the subgroup $U\x G\sub G\x G$, here re\-pre\-sen\-ta\-tion of 
the internal-external operation  group of the interactions.

A group  re\-pre\-sen\-ta\-tion on $U\x U\lq G$ 
involves subgroup $U$-intertwiners. 
The  $U$-intertwiners ${\bl B}$ of a group $G$ and a vector space $W$
with $U$-re\-pre\-sen\-ta\-tion $D:U\map\GL(W)$ 
are  group orbits $G\map W$, compatible with the $U$-action (taken as left action)
\begin{eq}{l}
\Diagr GGWW {L_u}{\bl B} {D(u)}{{\bl B}},~~\begin{array}{l}
u\in U,~g\in G\cr
\bl B(g)=D(u){\bl B}(u^{-1}g)\end{array}
\end{eq}The intertwiners constitute the  $U$-invariant  subspace 
$W^{U\lq G}$ in the vector space $W^G$ with all mappings.
The dimension of 
 $W^{U\lq G}$ is the cardinality of
the sym\-met\-ric space $U\lq G$ times the dimension of the $U$-re\-pre\-sen\-ta\-tion space
$W$. The intertwiners can be looked at to be mappings on the classes 
$U\lq G$, i.e. from the subgroup $U$-orbits in 
the  group $G$ into the $U$-orbits in the vector space $W$
\begin{eq}{l}
 U\lq G\map V/U,~~Ug\mape {\bl B}(Ug)=D[U]{\bl B}(g) 
\end{eq}With  a $G$-invariant measure 
$d\mu(Ug)$ of the $U$-orbits the
intertwiners can be expanded as direct integral
with the values on the classes 
\begin{eq}{l} 
\bl B
=\plint_{U\lq G} d\mu (Ug) ~\bl B(Ug)
=\plint_{U\lq G} d\mu (Ug) ~\ro e(Ug)^a\bl B(Ug)_a 
\end{eq}Here $\{\ro e^a\}$ is a basis of $W$,
$\{\ro e(Ug)^a\}$ a basis distribution of the re\-pre\-sen\-ta\-tion space 
at each  class $W(Ug)=W\x\{Ug\}\cong W$
and $\bl B(Ug)_a$ the corresponding coefficients. 

The $G$-re\-pre\-sen\-ta\-tion, induced  from the $U$-re\-pre\-sen\-ta\-tion on $W$,
is the right action  with the full group $G$
 on the $U$-intertwiners 
\begin{eq}{l}
k\in G:~~ W^{U\lq G}\map W^{U\lq G},~~ 
\bl B\mape \bl B_k=\plint_{U\lq G} d\mu (Ug) ~\ro e(Ug)^a\bl B(Ugk)_a
\end{eq}The right $G$-action $U\lq G\x G\map U\lq G$ has $U$-isomorphic
fixgroups.

In general, the induced $G$-re\-pre\-sen\-ta\-tion on the intertwiner vector space
$W^{U\lq G}$
is  highly decomposable
\begin{eq}{rl}
W^{U\lq G}&={\PL_I}W^I, ~~I\in\rep G\cr
\bl B &={\PL_I}~
\plint_{U\lq G} d\mu (Ug) ~\bl B(Ug)^I
={\PL_I}~\plint_{U\lq G} d\mu (Ug) ~\ro e(Ug)^a\bl B(Ug)^I_a
\end{eq}For compact groups
$G\sup U$ one has Frobenius' theorem\cite{FULHAR,FOL} for the multiplicity 
$n(D_G)$ of
an irreducible $G$-re\-pre\-sen\-ta\-tion $D_G$, 
induced by an irreducible $U$-re\-pre\-sen\-ta\-tion $D_U$ - given by $
n(D_G)=n(D_G/D_U)$, i.e. the multiplicity of $D_U$ in $D_G$.
  
The expansion coefficients $\bl B(Ug)^I_a$, called transmutators
\cite{S003,S011},
constitute  - for finite dimensional $G$ and
$U$-representations -  rectangular matrices, e.g. the $(1\x 3)$-matrices
in the chapter introduction. They have  a typical hybrid left-right
transformation behavior for the product group $U\x  G$.

\subsection{Free Particle Fields as Spin Intertwiners}

Familiar examples for intertwiners and the inducing procedure are
the  particle fields. 
Particle fields use the inducing procedure 
for the Poincar\'e group via the Lorentz group action on the energy-momenta
with the fixgroups (`little groups')
$\{\SO_0(1,3),\SO_0(1,2),\SO(3),\SO(2)\sx\R^2\}$.

The re\-pre\-sen\-ta\-tions of the
Poincare group $G=\SL(\C^2)\sx\R^4$ are induced, for massive particles, 
from $U=\SU(2)\x \R^4$-re\-pre\-sen\-ta\-tions. The inducing procedure  involves
the spin classes in the Lorentz group
\begin{eq}{l}
\SL(\C^2)=\SL(\C^2)/\SU(2)\x\SU(2)
\end{eq}Particle fields are spin intertwiners and can be written  
as a direct integral with Lorentz invariant boost measure
${d^3q\over q_0}$,
 pa\-ra\-met\-ri\-zed by the momenta 
${\rvec q\over m}\in\SL(\C^2)/\SU(2)\cong\R^3$
 of the mass hyperboloid $q^2=m^2$.
Creation and annihilation operators give a basis distribution for the spin $J$ 
re\-pre\-sen\-ta\-tion spaces
$W(\rvec q )\cong\C^{1+2J}\x \{\rvec q\}$ for each momentum, e.g.
the electron-positron
\begin{eq}{rl}
\bl\Psi&=
{\scriptsize\pmatrix{
 \bl l^{\dot A}\cr
\bl  r^{ A}\cr}}= \sqrt{2m}\plintq3
{\scriptsize\pmatrix{s({q\over m})^{\dot A}_a
 {\ro u(\rvec q)^a+\ro a^\star(\rvec q)^ a\over\sqrt2}\cr
\hat s({q\over m})_a^{ A}
{ \ro u(\rvec q)^a- \ro a^\star(\rvec q)^a\over\sqrt2}\cr}}\hbox{ with }
q_0=\sqrt{m^2+\rvec q^2}\cr
\end{eq}The expansion coefficients (spin-Lorentz group transmutators)
are the fundamental left and right  handed  
$(2\x2)$-Weyl boost re\-pre\-sen\-ta\-tions 
\begin{eq}{l}
s({q\over m})_a^{\dot A}\cong
\sqrt{{m+q_0\over 2m}}(\bl 1_2
+{\rvec q\over m+q_0})
 \hbox{ with }\left\{
\begin{array}{rrl}
\hbox{spin }\SU(2):&a&=1,2\cr
\hbox{Lorentz }\SL(\C^2):&\dot A&=1,2\cr
\end{array}\right.\cr
\end{eq}The indices are of `different quality' - for  a vector space
$V\cong\C^2$
with Lorentz group Weyl re\-pre\-sen\-ta\-tion and for the inducing vector space 
$W\cong\C^2$ with spin $J={1\over2 }$ re\-pre\-sen\-ta\-tion.

With  the action of the  product group
$\SL(\C^2)\x\SU(2)$ there are two different transformation 
structures: The  transmutators,
pa\-ra\-met\-ri\-zed with momenta,  are acted upon  with the
extended Lorentz group (from left). This produces the 
boosts for the momenta  in $q\mape \la q\la^\star$
and the Wigner $\SU(2)$-rotation (from right), dependent on the acting $\la$
and the class ${\rvec q\over m}$  
\begin{eq}{l}
\la\in\SL(\C^2):~~\la\o  s({q\over m})=s(\la{q\over m}\la^\star)\o
r(\la,{q\over m})
,~~r(\la,{q\over m})\in\SU(2)
\end{eq}Under Lorentz group action
the creation-annihilation operators transform with the corresponding Wigner 
$\SU(2)$-rotation
\begin{eq}{rl}
\la\in\SL(\C^2):~~
\ro u(\la\rvec q\la^*)^a&=r(\la,{q\over m})^a_b \ro u(\rvec q)^b\cr 
\then \bl l^{\dot A}&\mape \la^{\dot A}_{\dot B} \bl l^{\dot B}
\end{eq}

Products of  the fundamental boost re\-pre\-sen\-ta\-tions
give all 
$\SL(\C^2)/\SU(2)$ transmutators, e.g. the 
rectangular $(1\x4)$ and $(3\x 4)$
re\-pre\-sen\-ta\-tions in $\SO_0(1,3)/\SO(3)$
by the 1st and the last three columns of the $(4\x4)$-matrix
\begin{eq}{l}
\La({{q\over m}})^j_k\cong{1\over2}
\tr s({q\over m})\si^j s^\star({q\over m})\d\si_k,~~
\La({ {q\over m}})
={\scriptsize\left(\begin{array}{c|c}
{q_0\over m}&{q_a\over m}\cr
{q_b\over m}&\de_{ab}+{q_aq_b\over m(m+q_0)}\cr\end{array}\right)}
\end{eq}which are used to  
induce Poincar\'e group re\-pre\-sen\-ta\-tions for 
particles with spin $J=0$ (massive scalar particles)  and $J=1$ 
(massive vector particles) resp. This is in analogy to the
column decomposition of the $\SO(3)$-example 
in the chapter introduction.

\subsection{Interaction Fields as Hyperisospin Interwiners}  

Interaction fields   use the inducing  procedure for
the extended Lorentz group action 
arising  from hyperisospin re\-pre\-sen\-ta\-tions.

Hyperisospin interwiners ${\bl B}$
relate nonlinear spacetime
$x_\od\in\D(2)\cong\U(2)\lq\GL(\C^2)$, 
pa\-ra\-met\-ri\-zing the hyperisospin orbits $\U(2)g$
in the polar decomposition $\GL(\C^2)=\U(2)\x \D(2)$,
 to hyperisospin $\U(2)$-orbits in a re\-pre\-sen\-ta\-tion space
 $W\cong\C^n$.
They can be expanded with a direct integral over the future cone     
\begin{eq}{l}
\bl B=\plint_{\R^4_\od}{d^4 x\over(x^2)^2}~ \ro e(x)^\al\bl B(x)_\al\cr
\end{eq}with  invariant measure of nonlinear spacetime
\begin{eq}{l}
\R^4_\od\ni x=e^{{\psi_0}  +\rvec\psi}:~{d^4 x\over (x^2)^2}
= d{\psi_0}  ~\sinh^2\psi~ d\psi~d\Om^2 
= d{\psi_0}  ~\psi^2 d\psi~d\Om^2+\dots\cr 
\end{eq}$\{\ro e(x)^\al\}$ is a basis of the space 
$W(x)$ with $\U(2)$-re\-pre\-sen\-ta\-tion at each
point of the forward cone.

The induced re\-pre\-sen\-ta\-tion
 on the hyperisospin intertwiner vector space
\begin{eq}{l}
\bl B\in W^{\R^4_\od}=
\plint_{\R^4_\od} {d^4x\over (x^2)^2}~W(x),~~W(x)\cong W
\end{eq}is  decomposable. 
A decomposition 
into nondecomposable $\GL(\C^2)$-re\-pre\-sen\-ta\-tion spaces
contains only such  Lorentz group $\SL(\C^2)$-re\-pre\-sen\-ta\-tions
which have the inducing isospin  $\SU(2)$-re\-pre\-sen\-ta\-tion   as 
spin $\SU(2)$-subre\-pre\-sen\-ta\-tion.
E.g. an internal  $\U(2)$-re\-pre\-sen\-ta\-tion with  isopin doublets (triplets)
induces only external $\GL(\C^2)$-re\-pre\-sen\-ta\-tions involving
spin doublets (triplets). 

This re\-pre\-sen\-ta\-tion inclusion property 
relating external and internal interaction
sym\-met\-ries can be checked: 
Indeed, with the exception of the Higgs field, all interaction pa\-ra\-met\-ri\-zing 
fields of the minimal
standard model have this property. The isospin $\SU(2)$-re\-pre\-sen\-ta\-tions 
are denoted by their invariant  $T$ 
\begin{eq}{l}
[2T]:\SU(2)\map\SU(1+2T)\hbox{ with }T\in\{0,{1\over2},\dots\}\cr
\end{eq}and the Lorentz group 
$\SL(\C^2)$-re\-pre\-sen\-ta\-tions 
by their left and right-handed invariants  $J_L,J_R$
\begin{eq}{l}
[2J_L|2J_R]:\SL(\C^2)\map\SL(\C^{(1+2J_L)(1+2J_R)})
\hbox{ with }J_L,J_R\in\{0,{1\over2},\dots\}\cr
\end{eq}The Lorentz group re\-pre\-sen\-ta\-tions are  spin
$\SU(2)$-decomposable  
\begin{eq}{l}
[2J_L|2J_R]\stackrel{\SU(2)}\cong {\PL_{J=|J_L-J_R|}^{J_L+J_R}}[2J]
\end{eq}The standard model fields,
left and right handed fermions and gauge fields, 
are acted upon with the following isospin-spin
re\-presentations
\begin{eq}{c}
{\scriptsize\begin{array}{|c||c|c|c|c|}\hline
\hbox{interaction field}&\hbox{isospin}&\hbox{Lorentz group}&
\hbox{hypercharge}&\hbox{color}
\cr
&\SU(2),~[2T]&
\SL(\C^2),~[2J_L|2J_R]&\U(1),~Y&\SU(3),~[2C_1,2C_2]\cr\hline\hline
\hbox{left leptons }\bl l(x)&[1]&[1|0]&-{1\over2}&[0,0]\cr
\hbox{right lepton }\bl r(x)&[0]&[0|1]
&-1&[0,0]\cr
\hbox{left quarks }\bl q(x)&[1]&[1|0]&{1\over6}&[1,0]\cr
\hbox{right quarks }\bl u(x),\bl d(x)&[0]&[0|1]
&{2\over3},-{1\over3}&[1,0]\cr\hline
\hbox{hypercharge gauge }\bl A_0(x)&[0]&[1|1]&0&[0,0]\cr
\hbox{isospin gauge  }\rvec{\bl A}(x)&[2]&[1|1]&0&[0,0]\cr
\hbox{color gauge  }\bl g(x)&[0]&[1|1]&0&[1,1]\cr\hline\hline
\hbox{Higgs  } h(x)&[1]&[0|0]&{1\over2}&[0,0]\cr\hline

\end{array}}\cr\cr

\hbox{\bf isospin induced 
spin re\-pre\-sen\-ta\-tions}\cr
\hbox{$\SU(2)$ induces $\SL(\C^2)$ with }T\le J_{\rm max}=J_L+J_R
\end{eq}With the exception of the Higgs field,
all isospin re\-pre\-sen\-ta\-tions for 
the interaction fields are Lorentz group subre\-pre\-sen\-ta\-tions.
With $\SU(3)$ 
 no subgroup of $\GL(\C^2)$,
the color degrees of freedom 
are not involved in the Lorentz group re\-pre\-sen\-ta\-tion 
inducing procedure. The special role of the Higgs degrees of freedom will be
discussed in the next section.

\section{The Higgs Hilbert Space}

Inducing a $G$-re\-pre\-sen\-ta\-tion from a subgroup $U$-re\-pre\-sen\-ta\-tion on
a Hilbert space $W\cong\C^n$
\begin{eq}{l}
W\x W\map\C,~~\sprod {\bl B}{\bl B'}\hbox{ with } \sprod {\bl B}{\bl B}>0
\hbox{ and }\sprod
{\bl B}{\bl B}=0\iff \bl B=0\cr
\hbox{Hilbert basis: }\sprod {\ro e^a}{\ro e^b}=\de^{ab},~~\bl B=\ro e^a\bl B_a,
~~\sprod {\bl B}{\bl B'}=\ol{\bl B_a}\bl B'_a
\end{eq}the intertwiners inherit, 
with a positive invariant
$U\lq G$ measure $d\mu(Uk)$,  a scalar product
\begin{eq}{rl}
\bl B
&=\plint_{U\lq G} d\mu (Ug) ~ \ro e(x)^\al\bl B(Ug)_a\cr
W^{U\lq G}\x W^{U\lq G}\map\C,~
\sprod{\bl B}{\bl B}&=\int_{U\lq G} d\mu (Ug) ~
\ol{\bl B(Ug)_a}\bl B'(Ug)_a 
\end{eq}In contrast to compact groups, the scalar product
for a noncompact group $G$ has not to be finite if restricted to
 $G$-subre\-pre\-sen\-ta\-tions $W^I\subnoteq W^{U\lq G}$. E.g., 
the Weyl boost re\-pre\-sen\-ta\-tions $s({q\over m})$ are not 
${d^3q\over q_0}$-square integrable - momentum  wave packets have to be used.

The basis distributions $\{\ro e(Ug)^a\}$
have `continuous orthonormality'
\begin{eq}{l}
\sprod{\ro e(Ug)^a}{\ro e(Ug')^b}=\de^{ab}\de(\mu(Ug,Ug'))\cr
\end{eq}with 
a Dirac distribution, associated  to
the  $U\lq G$-measure 
\begin{eq}{rl}
\int_{U\lq G\x U\lq G} 
d\mu (Ug)d\mu (Ug') ~f(Ug,Ug')\de(\mu(Ug,Ug'))
=\int_{U\lq G} 
d\mu (Ug) ~f(Ug,Ug)
\end{eq}

The particle fields above are examples with 
the spin Hilbert spaces $W\cong\C^{1+2J}$, e.g. the 
electron-positron field
\begin{eq}{rl}
 \bl l^{\dot A}
 &= \sqrt{2m}\plintq3
s({q\over m})^{\dot A}_a
 {\ro u(\rvec q)^a+\ro a^\star(\rvec q)^ a\over\sqrt2}
\hbox{ with }
 q_0=\sqrt{m^2+\rvec q^2}\cr
\SL(\C^2)/\SU(2)&\cong\cl Y^3\cong \R^3\ni {\rvec q\over m}\mape\ro u(\rvec q)^a\in W\cong \C^2\cr
&\sprod{\ro u(\rvec q)^a}{\ro u(\rvec q')^b}=
\de^{ab} 2q_0\de({\rvec q-\rvec q'\over 2\pi})\cr
\end{eq}

What the momentum dependent 
creation-an\-ni\-hi\-la\-tion operators
are  for the induced Poincar\'e group representation, that
is the spacetime dependent Higgs field which
induces extended Lorentz group $\GL(\C^2)$-re\-pre\-sen\-ta\-tions from 
hyperisospin $\U(2)$-re\-pre\-sen\-ta\-tions.
The  Higgs field is a basis  distribution for the Hilbert space  $H\cong\C^2$ 
 with the defining $\U(2)$-re\-pre\-sen\-ta\-tion 
\begin{eq}{rl}
\bl B&=\plint_{\R_\od^4} {d^4 x\over(x^2)^2}
~~ h(x)^\al\bl B(x)_\al\cr
\U(2)\lq \GL(\C^2)&\cong\R^4_\od\ni x\mape  h(x)^\al\in H\cong \C^2\cr
&\sprod{ h(x)^\al}{ h(x')^\be}=\de^{\al\be}(x^2)^2\vth(z^2)\de(z),~~z=x-x'
\end{eq}E.g., the left lepton fields can arise as an irreducible 
$\GL(\C^2)$-re\-pre\-sen\-ta\-tion induced by the Higgs field
\begin{eq}{l}
\bl l^{\dot A}=\plint_{\R_\od^4} {d^4 x\over(x^2)^2}
~ h(x)^\al\bl l(x)^{\dot A}_\al 
 \hbox{ with }\left\{
\begin{array}{rrl}
\hbox{Lorentz }\SL(\C^2):&\dot A&=1,2\cr
\hbox{isospin }\SU(2):&\al&=1,2\cr
\end{array}\right.\cr
\end{eq}$\bl l^{\dot A}$ has two expansions:
A Higgs expansion  for  
nonlinear spacetime $x\in \R^4_\od$ as
interaction field with hyperisospin action,
 and a creation-annihilation expansion for
its momentum hyperboloid ${\rvec q\over m}\in\cl Y^3$   
as particle field with Poincar\'e group action.

\section{A  Fermi-Clifford Algebra for Hyperisospin} 

In the last section an attempt   will be described to
induce all standard model interaction fields from
$\U(2)$-re\-pre\-sen\-ta\-tions on the  Higgs degrees of freedom.

The hypercharges of the interaction fields in the standard model
\begin{eq}{l}
Y\in\{0,\pm {1\over6},\pm {1\over3},\pm {1\over2},\pm {2\over3},\pm 1\}
\end{eq}suggest the grading structure of a Grassmann
algebra whose basic vector space is
complex 2-dimensional and will be identified with the Higgs Hilbert space
$H\cong\C^2$ and its dual $H^\star\cong\C^2$
\begin{eq}{l}
\U(2)\x (H\pl H^\star)\map H\pl H^\star,~~\left\{\begin{array}{rl}
u.  H&=e^{i\al_0+i\rvec\al} H\cr
u.  H^\star&= H^\star e^{-i\al_0-i\rvec\al}\cr\end{array}\right.\cr
\hbox{hypercharge-isospin invariants: }
[2Y;2T]=\left\{\begin{array}{rl}
[1;1]&\hbox{for }H\cr
[-1;1]&\hbox{for }H^\star\cr\end{array}\right.\cr
\hbox{bases for $H$, $H^\star$: }\{ h^\al,h^\star_\al\mid\al=1,2\}
\end{eq}

The Grassmann algebra   $\And[H\pl H^\star]$  of Higgs and anti-Higgs 
Hilbert space is the
direct sum of their  totally antisym\-met\-rized
tensor products. It 
has complex dimension $2^4$ and 
comes with  a $\Z_5$-graduation,
given by the hypercharge invariants $2Y\in \Z_5=\{0,\pm 1,\pm 2\}$.
The ${4\choose 2-2|Y|}$-dimensional spaces with grade $2Y$ are
decomposable into
irreducible isospin re\-pre\-sen\-ta\-tion spaces with dimension $1+2T$
\begin{eq}{c}
\begin{array}{|l||c|c|c|}\hline
&2T=0&2T=1&2T=2\cr\hline\hline
2Y=+2& h^\al\ep_{\al\be} h^\be&-&-\cr\hline
2Y=-2& h^\star_\al\ep^{\al\be} h^\star_\be&-&-\cr\hline
2Y=+1&-& h^\al,~~ h^\star_\al h^\ga\ep_{\ga\be} h^\be
&-\cr\hline
2Y=-1&-& h^\star_\al,~~ h^\al h^\star_\ga\ep^{\ga\be} h^\star_\be&-\cr\hline
2Y=0&1,~ h^\star_\al h^\al,~
( h^\star_\al h^\al)^2&-& h^\star_\al\rvec \tau^\al_\be h^\be\cr\hline
\end{array}\cr
\cr
\hbox{\bf hyperisospin properties of the Higgs-Grassmann algebra}\cr

\AND(H\pl H^\star)\cong\C^{16}
\end{eq}

The Grassmann vector space is endowed with a Fermi structure\cite{S90} 
by nontrivial anticommutators for the
generating Higgs vectors
\begin{eq}{rl}
\hbox{Fermi anticommutators:}&
\acom{ h^\star_\al}{ h^\be}=\de^\be _\al,~~
\acom{ h^\al}{ h^\be}=0,~~
\acom{ h^\star_\al}{ h^\star_\be}=0
\end{eq}Bose and Fermi elements
have even and odd grades resp.
in the Fermi quantum algebra $\bl Q_+(\C^2)\cong\C^{16}$
which is the vector space structure of the Grassmann algebra with
nontrivial basic  anticommutators\cite{S922}.

The Fermi quantum algebra is isomorphic to
a complexified Clifford algebra with neutral signature
$\O(2,2)$ 
\begin{eq}{l}
\AND(H\pl H^\star)\cong
\bl Q_+(\C^2)\cong \C\ox\CLIFF(2,2),~~\CLIFF(2,2)\cong\R^{16}
\end{eq}This can be seen with a basis,
(anti)-hermitian 
with respect to the 
indefinite unitary conjugation, compatible with isopin $\SU(2)$
\begin{eq}{l}
 h^\be\stackrel{\star}\lrmap \ep^{\be\de} h^\star_\de:~~
\left\{
\begin{array}{ll}
\acom{{ h^\star_\al+ h^\ga\ep_{\ga\al}\over\sqrt2}}
{{ h^\be+\ep^{\be\de} h^\star_\de\over\sqrt2}}
&=\de_\al^\be\cr
\acom{{ h^\star_\al- h^\ga\ep_{\ga\al}\over i\sqrt2}}
{{ h^\be-\ep^{\be\de} h^\star_\de\over i\sqrt2}}
&=-\de_\al^\be\cr
\acom{{ h^\star_\al- h^\ga\ep_{\ga\al}\over i\sqrt2}}
{{ h^\be+\ep^{\be\de} h^\star_\de\over\sqrt2}}
&=0\end{array}\right.
\end{eq}$\CLIFF(2,2)$ is isomorphic to the real
$(4\x4)$-matrices\cite{LOUN,PORT,S013}.

The $\Z_5$-graduation with the 
power difference  of Higgs and
anti-Higgs space reflects the adjoint action of
the   hypercharge  $\U(1)$ generator 
\begin{eq}{l}
I=2Y={[ h^\star_\al, h^\al]\over 2}\then
[I,h^\al]=h^\al,~~[I,h^\star_\al]=-h^\star_\al
\end{eq}Isospin $\SU(2)$ has the generators
\begin{eq}{l}
\rvec T={ h^\star_\al\rvec\tau^\al_\be h^\be\over2}\then
[\rvec T,h^\al]={\rvec\tau^\al_\be\over2}h^\be,~~
[\rvec T,h^\star_\be]=-{\rvec\tau^\al_\be\over2}h^\star_\al
\end{eq}with the Casimir value given by the doubled adjoint action
$[\rvec T,[\rvec T,a]]=T(1+T)a$.

The hyperisopin re\-pre\-sen\-ta\-tions
can be related to and may induce 
 the extended Lorentz group $\GL(\C^2)$-re\-pre\-sen\-ta\-tions\cite{S982}
by expanding the standard model interaction fields 
with Higgs vector products as follows
\begin{eq}{c}

\begin{array}{|c||c|}\hline
\U(2)&\GL(\C^2)\cr\hline\hline
h^\al&\bl l_\al^{\dot A}\cr\hline
h^\star_\al h^\al&\bl A^0_k\cr\hline
h^\star_\al h^\be\rvec \tau^\al_\be&\rvec {\bl A}_k\cr\hline
h^\star_\be h^\ga h^\de\ep^{\al\be}\ep_{\ga\de}&
\bl Q_\al^{\dot A}\cr\hline
h^\star_\al h^\star_\be h^\ga h^\de\ep^{\al\be}\ep_{\ga\de}&
\bl  G\cr\hline
\end{array}\cr
\cr\hbox{\bf spacetime fields induced by the Higgs-Grassmann algebra}\cr
\hbox{e.g. }
\plint_{\R^4_\od}{d^4 x\over(x^2)^2}~
(h^\star h h)(x)^\al
\bl Q(x)_\al^{\dot A}
\end{eq}The standard model interaction fields are transmutators from
the hyperisospin $\U(2)$ group to the extended Lorentz group $\GL(\C^2)$. 
The Higgs field products involving more than one Higgs $h$ or more than one
anti-Higgs $h^\star$ induce  interaction fields which
can come with an additional  degree of freedom. 
E.g., the field  $\bl Q$, induced by a three Higgs's product, allows for its cubic root 
$\bl q$ an internal degree of freedom
whereof the product has to be a singlet, e.g., it is a 
left handed quark isodoublet $\bl q$
with  hypercharge color $\U(3)$ as cubic root of 
$\U(1)$
\begin{eq}{l}
\begin{array}{rllrl}
\bl q&=[\bl Q]^{1\over3},&\hbox{with}&\bl Q&=\bl q\and\bl q\and\bl q\cr
\U(3)&=[\U(1)]^{1\over3}&
\hbox{with}&\U(1)&=\U(3)\and\U(3)\and\U(3)\end{array}\cr
\plint_{\R^4_\od}{d^4 x\over(x^2)^2}~
(h^\star h h)(x)^\al
\bl Q(x)_\al^{\dot A}=
\plint_{\R^4_\od}{d^4 x\over(x^2)^2}~
(h^\star h h)(x)^\al
(\bl q\and\bl q\and\bl q)_\al^{\dot A}

\end{eq}A cubic root of a group can be defined via functions on the spectrum.
The new degrees of freedom arise in analogy to
the complex phases  in the cyclotomic groups 
$1^{1\over n}=\{e^{2\pi i{k\over n}}\mid k=1,\dots,n\}$.
Analogously, 
the square root  $\bl g=\sqrt{\bl G}$ of a four Higgs's product
allows the  Lorentz vector and color octet 
 re\-pre\-sen\-ta\-tion properties for a gluon field  $\bl g(x)_k^a$
 with $\bl G=\bl g^a_k\eta^{kl}\de_{ab}\bl g^b_l$.
  In such an approach,
color nontrivial fields for an effective approximation of the interaction
would be cubic and square roots of
fields without color. The hypercharge properties
of the color nontrivial fields match with such root 
properties\cite{HUCK,S921,S926,S941,S981}.

\end{document}